\documentclass[runningheads]{llncs}
\usepackage[english]{babel}
\usepackage[utf8]{inputenc}
\usepackage[T1]{fontenc}
\usepackage{lmodern}

\usepackage{xcolor}
 \usepackage{multirow}
\usepackage{tabularx}
\usepackage[ruled, vlined]{algorithm2e}
 \usepackage{subfigure}
\usepackage{graphicx}
\usepackage{amsmath,amsfonts,amssymb}
\usepackage{multicol}
\usepackage{psfrag}
\usepackage{pst-all}
\usepackage{url}
\usepackage{floatflt}
\usepackage{calc,pifont} \newcounter{local}
\renewcommand\theenumi{\protect\setcounter{local}%
{171+\the\value{enumi}}\protect\ding{\value{local}}}

\definecolor{dgreen}{rgb}{0,0.5,0}
\definecolor{dorange}{rgb}{0.8,0.4,0}
\definecolor{lblue}{rgb}{0.5,0.5,0.9}
\definecolor{dred}{rgb}{0.7,0.0,0.0}
\newcommand{\RefAlgo}[1]{Algorithm\,\ref{#1}}

\newcommand{\RefFigure}[1]{Fig.\,\ref{#1}}

\newcommand{\etal}{{\it et~al.}}

\begin{document}

\graphicspath{{Figs/}{./}}

\pagestyle{headings}
\mainmatter

\title{3D Geometric Analysis of Tubular Objects Based on Surface Normal Accumulation}

\titlerunning{3D Geometric Analysis of Tubular Objects }

 \author{
 	Bertrand Kerautret\inst{1,2} \and Adrien Krähenbühl\inst{1,2} \and Isabelle Debled-Rennesson \inst{1,2} \and Jacques-Olivier Lachaud \inst{3}
 	\thanks{This work was partially supported by the ANR grants DigitalSnow ANR-11-BS02-009}
 }
 \authorrunning{B. Kerautret \and A. Krähenbühl \and I. Debled-Rennesson \and J.-O. Lachaud}
 \institute{
 	Université de Lorraine, LORIA, UMR 7503, Vandoeuvre-lès-Nancy, F-54506, France
 	\and CNRS, LORIA, UMR 7503, Vandoeuvre-lès-Nancy, F-54506, France
 	\and LAMA (UMR CNRS 5127), Université Savoie Mont Blanc, F-73376, France
 	\email{\{bertrand.kerautret,adrien.krahenbuhl,isabelle.debled-rennesson\}@loria.fr}
 	\email{jacques-olivier.lachaud@univ-savoie.fr}
 }

\maketitle

\begin{abstract}
This paper proposes a simple and efficient method for the
reconstruction and extraction of geometric parameters from 3D tubular
objects. Our method constructs an image that accumulates surface
normal information, then peaks within this image are located by
tracking. Finally, the positions of these are optimized to lie
precisely on the tubular shape centerline. This method is very
versatile, and is able to process various input data types like full
or partial mesh acquired from 3D laser scans, 3D height map or
discrete volumetric images. The proposed algorithm is simple to
implement, contains few parameters and can be computed in linear time
with respect to the number of surface faces. Since the extracted tube
centerline is accurate, we are able to decompose the tube into
rectilinear parts and torus-like parts. This is done with a new linear
time 3D torus detection algorithm, which follows the same principle of
a previous work on 2D arc circle recognition. Detailed experiments
show the versatility, accuracy and robustness of our new method.
\end{abstract}

\section{Introduction}

Tubular shapes appear in various image application domains. They are
common in the medical imaging field. For instance, blood vessel
identification and measurements are an important object of study
\cite{kirbas2004,lesage_review_2009,tankyevych_angiographic_2011}. The wall thickness in bronchial
tree plays also an important role in several lung diseases
\cite{pare2012airway}. Tubular shapes also occur in CT volumetric
images of wood \cite{krahenbuhl2014}: their segmentation into knots is
exploited by agronomic researchers or in industrial sawmills. Outside
volumetric images, tubular objects are also present in industrial
context with the production of metallic pipes from bending
machines. Quality assessment of such metallic pieces is generally
achieved with a direct inspection by a laser scanner. Such process is
also performed for calibration purpose and reverse engineering tasks.

\begin{figure}
\center
\begin{tabular}{cccc}
\includegraphics[height=0.16\textwidth]{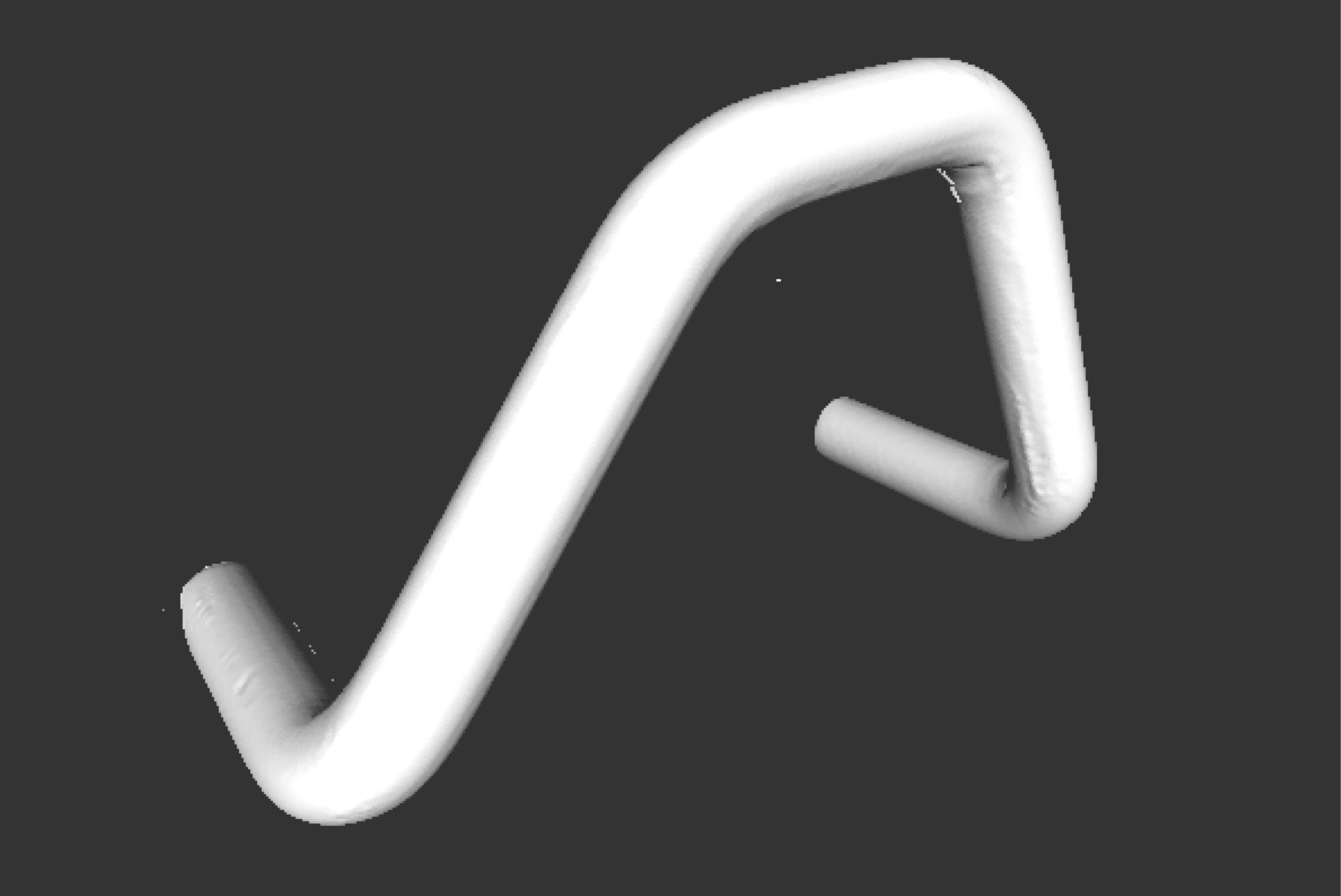}&
\includegraphics[height=0.16\textwidth]{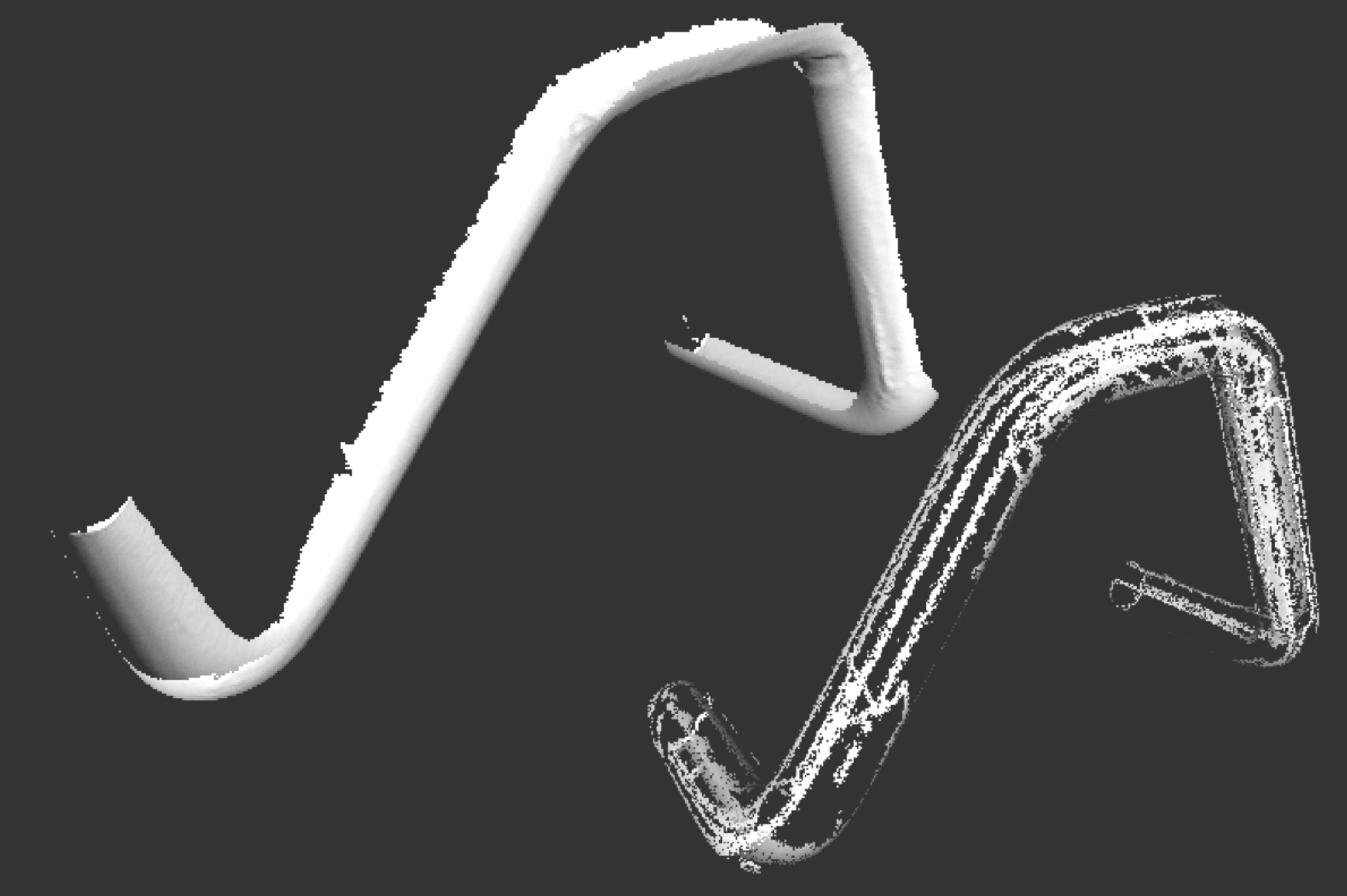}&
\includegraphics[height=0.16\textwidth]{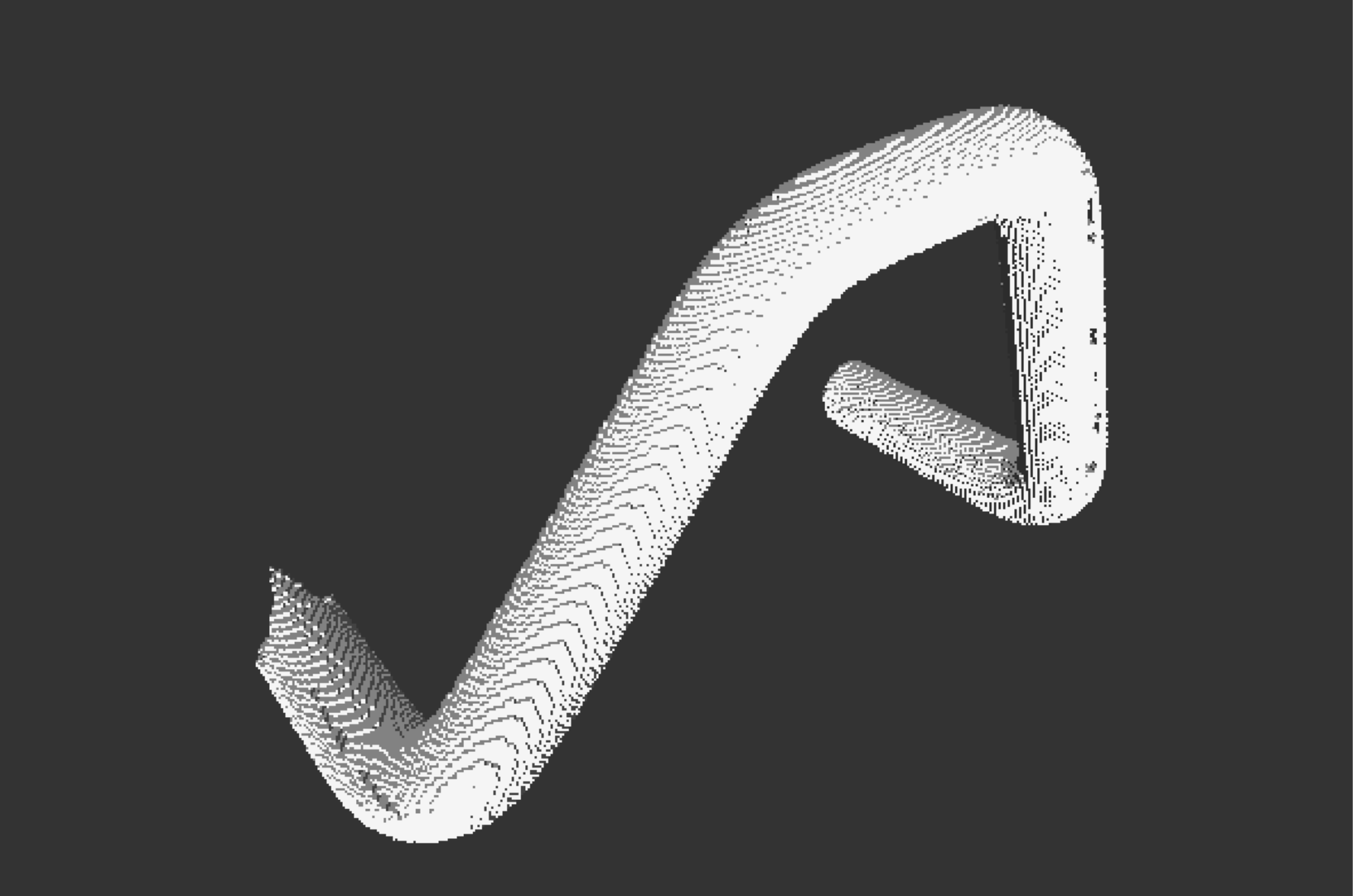}&
\includegraphics[height=0.16\textwidth, trim = 130 150 50 150, clip=true]{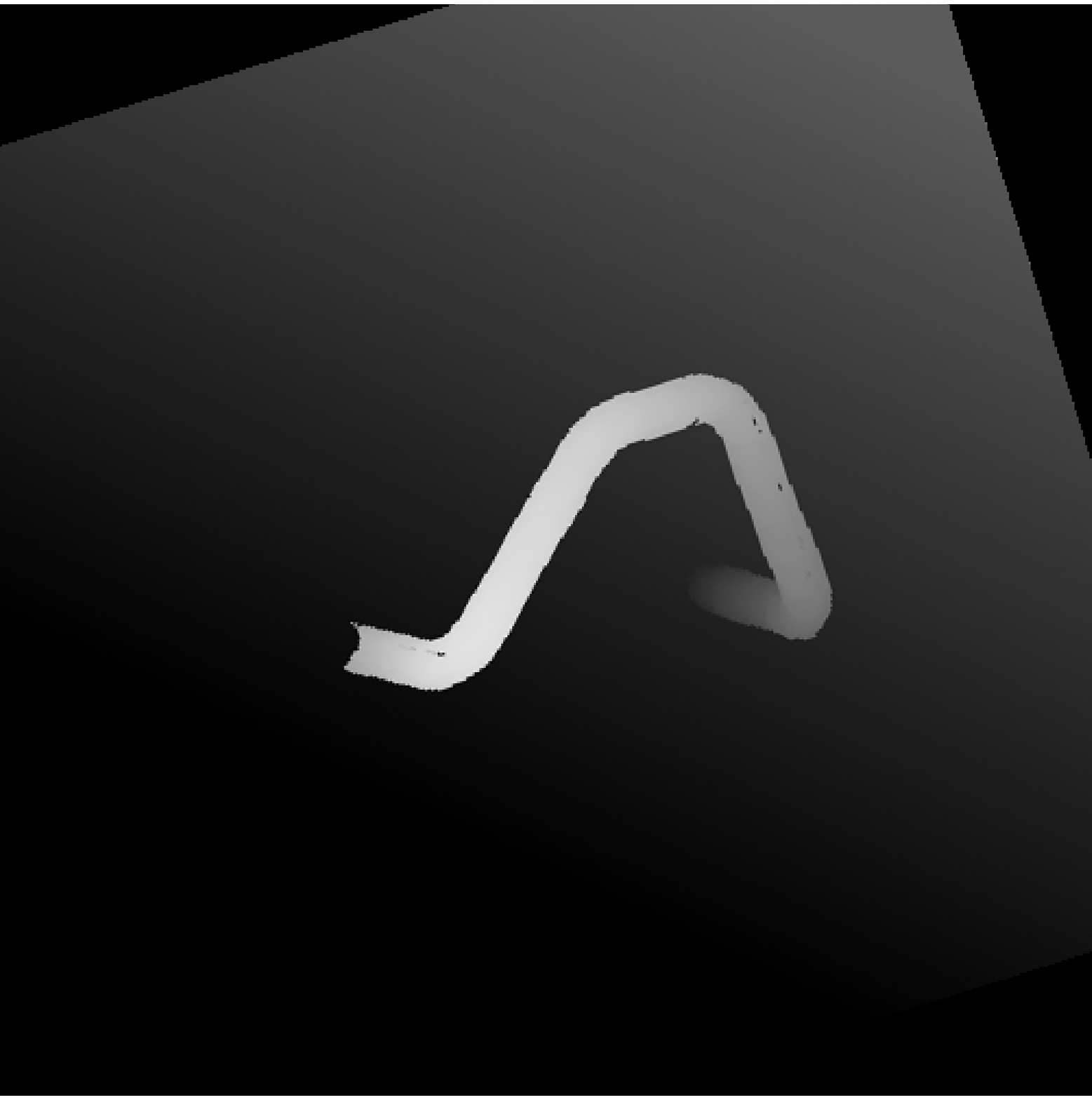}\\[-0.6cm]
\textcolor{white}{\textbf{(a)}} &\textcolor{white}{\hspace{-0.9cm}\textbf{ (b)} }&\textcolor{white}{\textbf{ (c)}} &\textcolor{white}{\textbf{ (d)}} 
\end{tabular}
\caption{Different kinds of 3D tubular data: (a) input data obtained
  from full laser scan, (b) partial scans from one direction, (c)
  digital set of voxels and (d) height map.}
\label{FigDataTypes}
\end{figure}

Geometric properties of tubular structures are extracted in different
ways depending on the application domain and on the nature of input
data. From unorganized set of points, Lee proposed a curve
reconstruction exploiting an Euclidean minimum spanning tree with a
thinning algorithm and applies it to pipe surface reconstruction
\cite{lee2000}. Later, Kim and Lee proposed another method based on
shrinking and moving least-squares \cite{lee2004} to improve the
reconstruction of pipes with non constant radius. However, as shown by
Bauer and Polthier \cite{bauer2009}, such a reconstruction method
produced noisy curves in particular for data extracted from partial
scans like the ones of \RefFigure{FigDataTypes}~(b). Another approach
estimates the principal curvatures of the set of points in order to
detect cylindrical and toric parts \cite{Goulette1997}. Although
promising, this approach suffers from the quality of the local
curvature estimator. To overcome this limitation, Bauer and Polthier
\cite{bauer2009} proposed to recover a parametric model based on a
tubular spine. Their method is able to process partial laser scans
limited to one particular direction. The main steps of their method
consists in first projecting the mesh points onto the spinal region of
the mesh before reconstructing a spine curve and analyzing it. The
method requires as parameter one radius size, and it cannot process
volumetric data (voxel sets) or heightmap data.

\begin{figure}
\center
\begin{tabular}{@{}c@{}c@{}c@{}c}
\includegraphics[width=0.23\textwidth, trim = 100 0 100 0, clip=true]{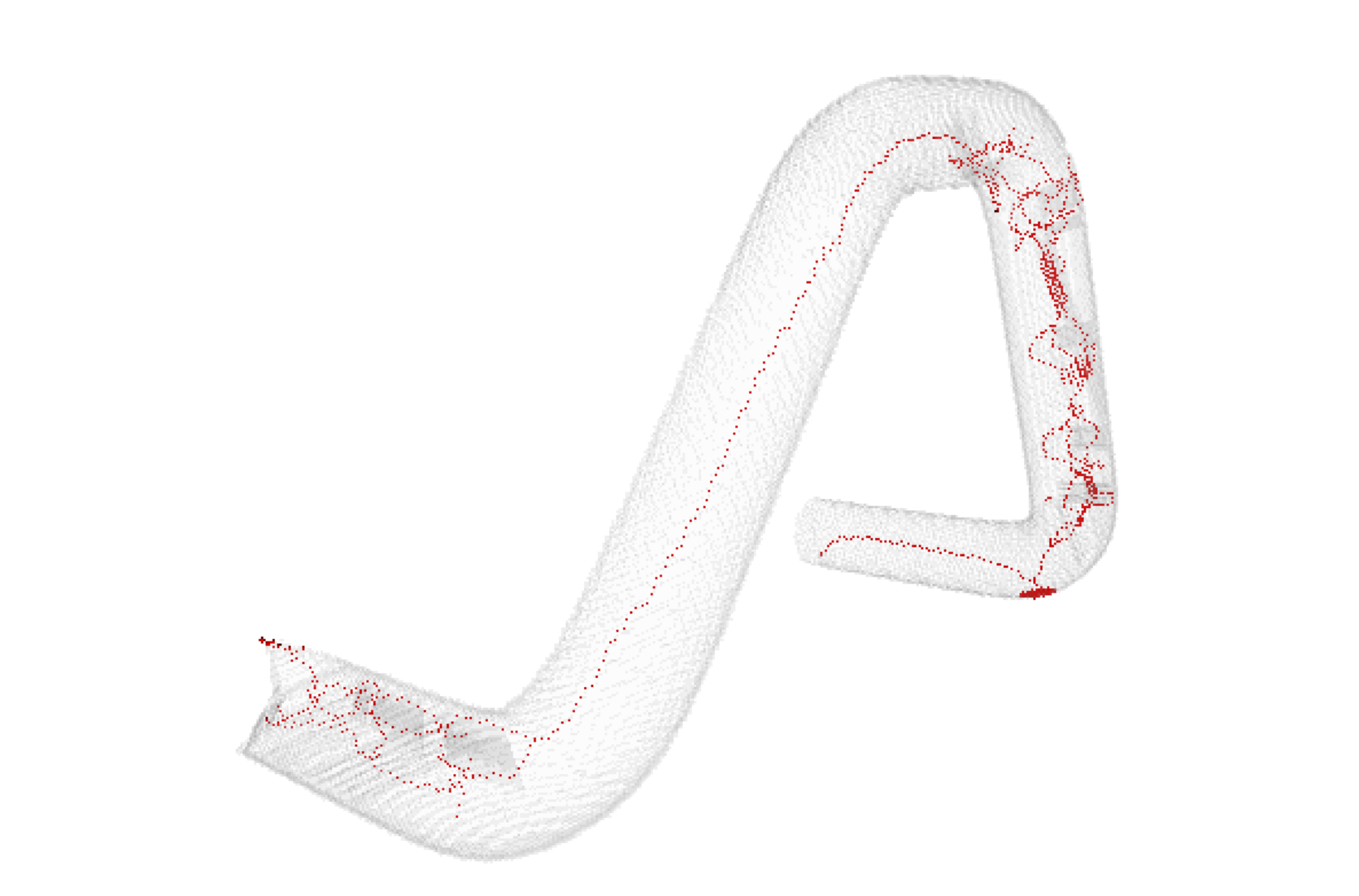}&
\includegraphics[width=0.23\textwidth, trim = 100 0 100 0, clip=true]{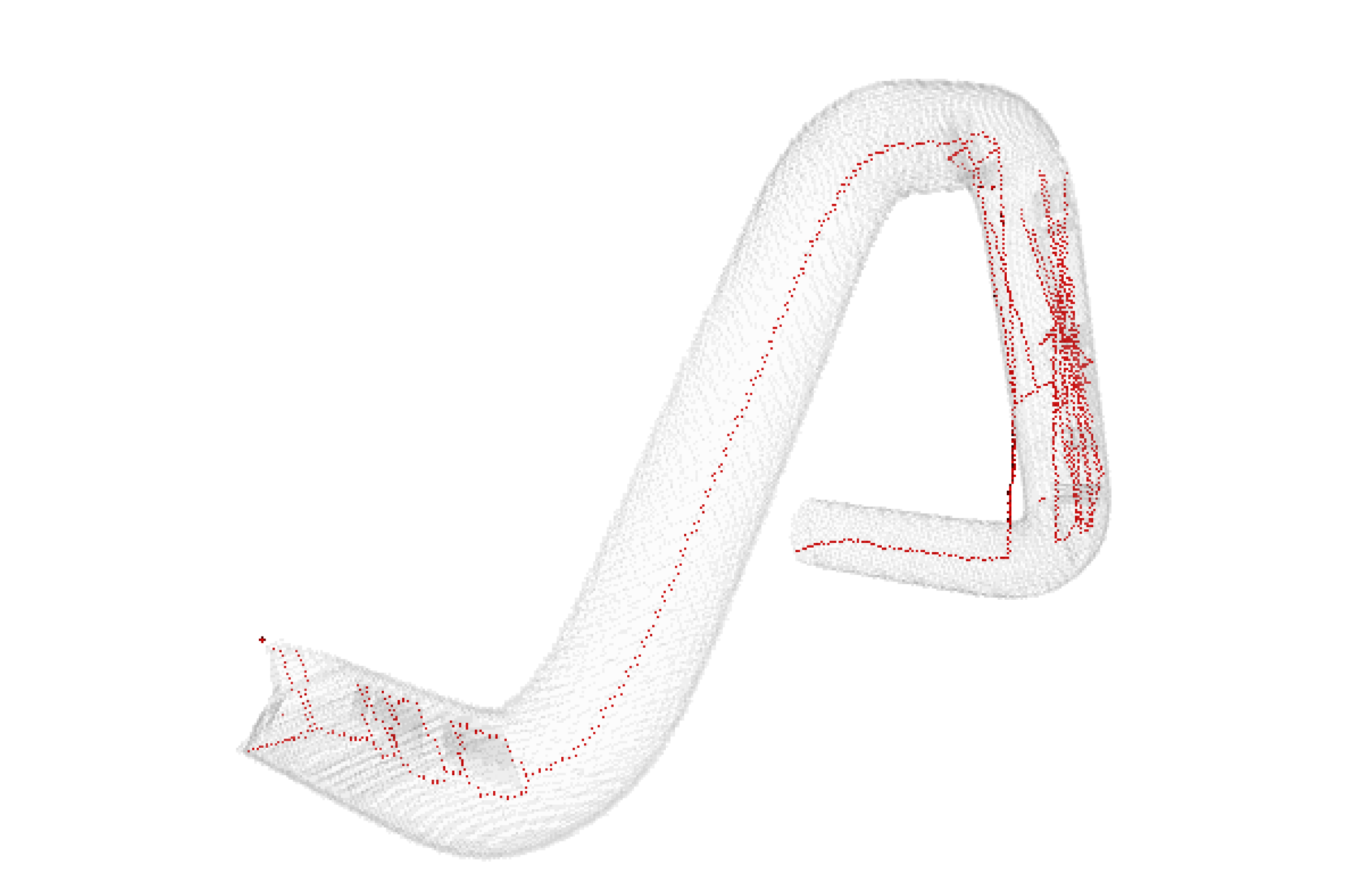}&
\includegraphics[width=0.23\textwidth, trim = 100 0 100 0, clip=true]{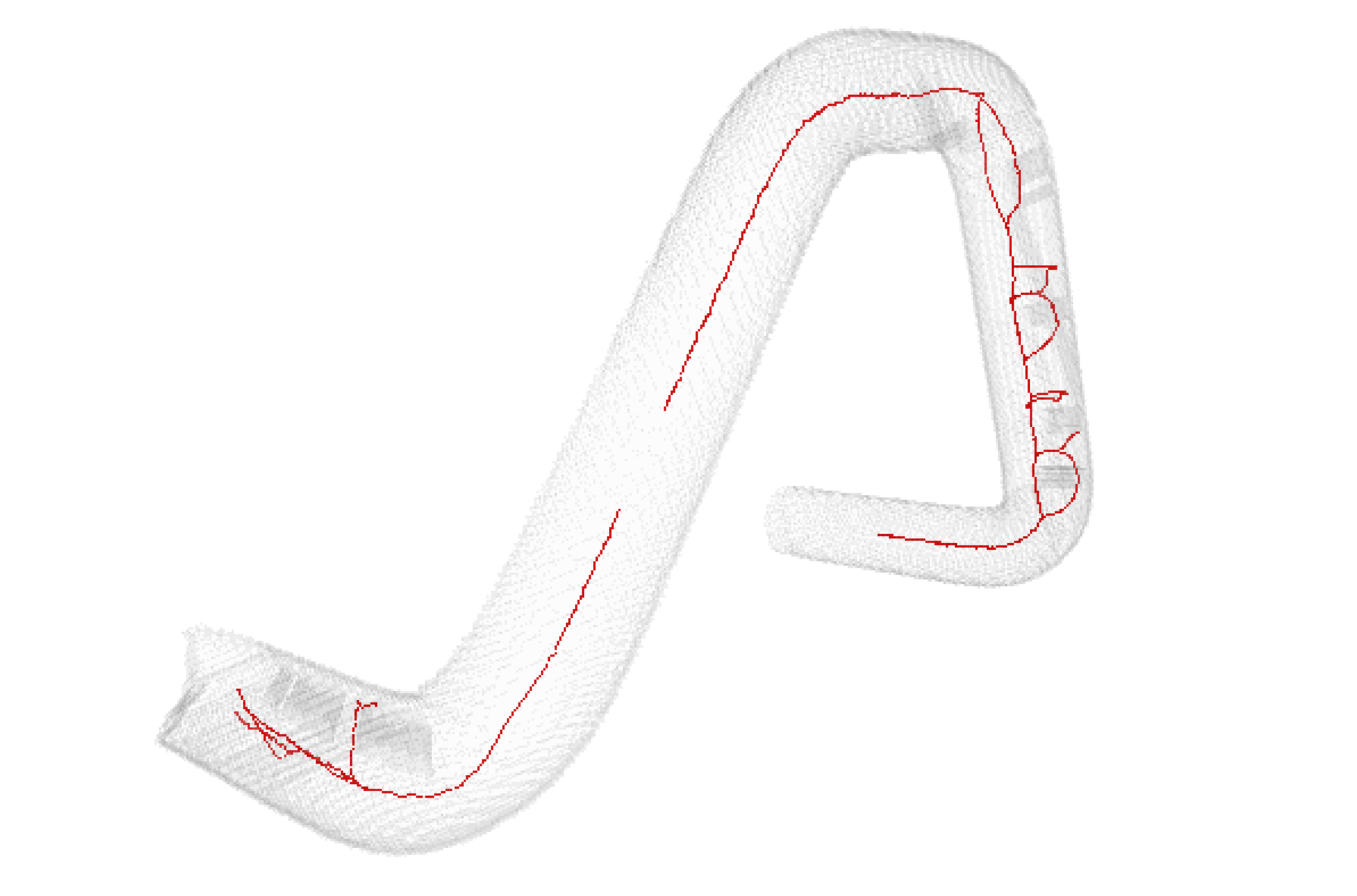}&
\includegraphics[width=0.23\textwidth, trim = 100 0 100 0, clip=true]{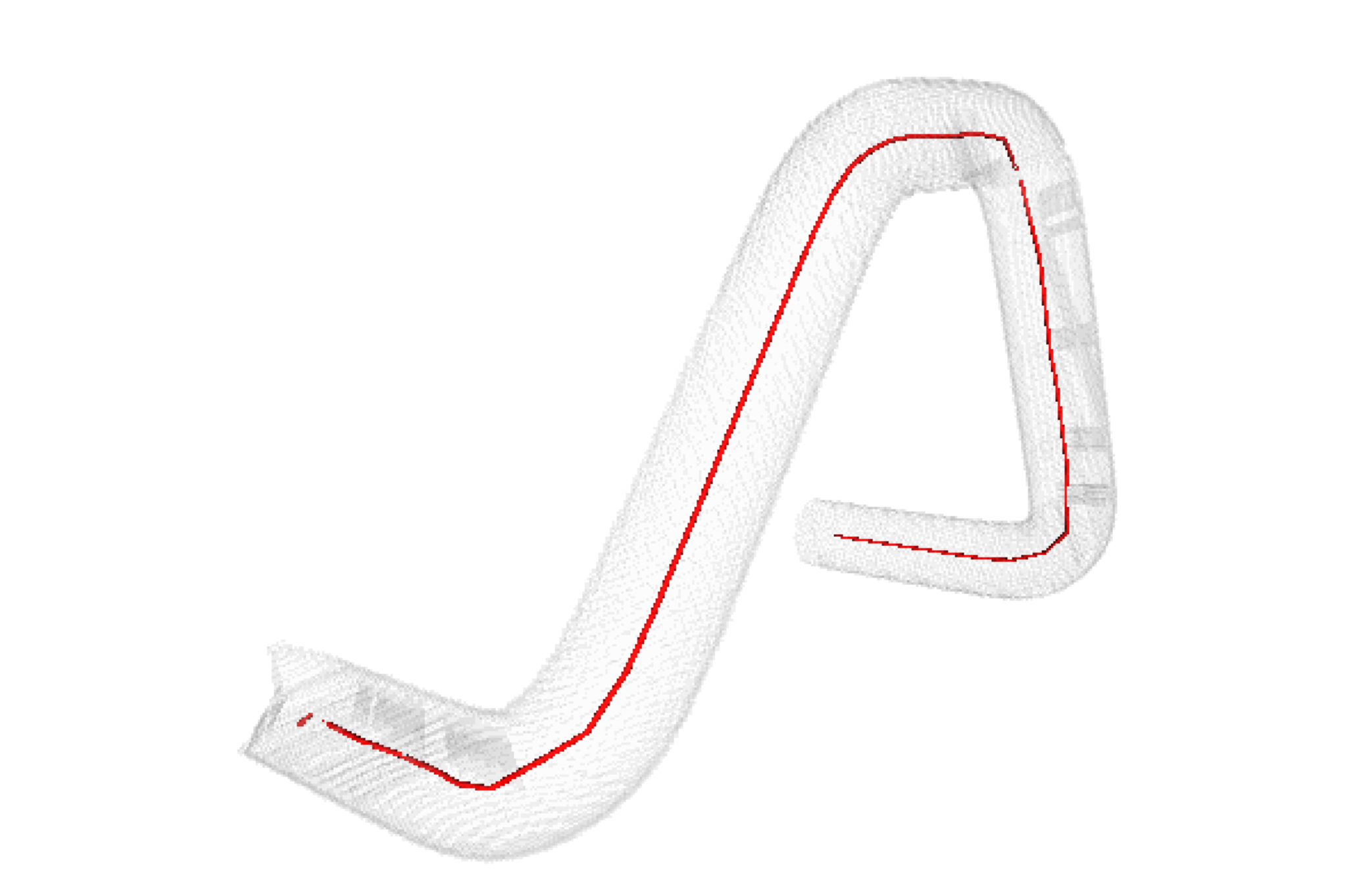}\\
{\small{(a) thinning}} &{\small{ (b) geometric}} & {\small{(c) potential field}} & {\small{(d) proposed }} \\
  {\small{1 min 4s}} &    {\small{0.5 s}} &  {\small{3min }} &   {\small{6s+3s}}
\end{tabular}
\caption{Skeleton extraction from three different methods presented in
  \cite{Cornea2007} with the implementation given by the authors. Our
  method is presented on the right.}
\label{FigComparisons}
\end{figure}

More generally, classic medial axis extraction looks to be a potential
solution for tubular shape analysis \cite{Cornea2007}. However such
extraction may be sensitive to noise or to the presence of small
defaults in the volumetric discrete object (like small
hole). \RefFigure{FigComparisons} shows some results obtained with
different methods available from the implementation given in the
authors survey. We can clearly see that small holes in the digital
object significantly degrade the result. More recently, many advances
came from the field of mesh processing with approaches based on mesh
contraction \cite{au2008,tagliasacchi2012}. However, they are
generally not adapted to surface with boundaries like the partial scan
data of \RefFigure{FigDataTypes}. In the same way, they are not simple
to adapt to volumetric data like digital object made of voxels or
height map. To process volumetric discrete objects, Gradient Vector
Flow \cite{xu_snakes_1998} was exploited by Hassouna and Farag in
order to propose a robust skeleton curve extraction
\cite{hassouna_variational_2009}. This method was also adapted to
process gray values volumetric images used in virtual endoscopy
\cite{bauer_extracting_2008}. In the field of discrete geometry we can
mention a method which propose to specifically exploit 3D discrete
tools to extract some medial axis on grey-level images
\cite{baja2005}.  To sum up approaches on medial axis, they are
designed to process shapes defined as volumes, but they fail when
processing open surfaces or partial samplings of the shape
boundary. To process such data, we can refer to the work of
Tagliasacchi {\etal} \cite{tagliasacchi_curve_2009}, who propose an algorithm based
on surface normals. However the resulting quality depends on manual
parameter tuning.

In this work, we propose a unified approach to the reconstruction and
the geometric analysis of tubular objects obtained from various input
data types: laser scans sampling the shape boundary with partial or
complete data (\RefFigure{FigDataTypes}~(a,b)), voxel sets sampling
the shape (\RefFigure{FigDataTypes}~(c)) or more specifically from
height map data (\RefFigure{FigDataTypes} (c,d)). Potential
applications of the latter datatype are numerous because of the
increasing development of \textit{Kinect}{\textregistered}-like
devices. Our main contributions are first to propose a simple and
automatic centerline extraction algorithm, which mainly relies on a
surface normal accumulation image. Like other Hough transform based
applications \cite{borrmann_3d_2011,tarsha-kurdi_hough-transform_2007}, this
algorithm can handle various types of input data. We also propose to
extract geometric information along the tubular object, by segmenting
it into rectilinear and toric parts. This is achieved with a 3D
extension of a previous work on circular arc detection along 2D
curves. In the following sections, we first introduce the new method
of centerline detection, then we show how to reconstruct the tubular
shape and decompose it into meaningful parts. We conclude with
representative experiments showing the qualities of our method.

\section{Fast and Simple Centerline Extraction on 3D Tubular Object}

In this section, we present centerline extraction algorithm based on
surface normal accumulation. It consists in three main steps. First,
we compute a 3D accumulation image, which counts for each voxel how
many faces of input data have their normal vector pointing through the
voxel (i). Depending on input data, normal vectors can be defined
directly from the mesh faces or estimated by a more robust and
accurate estimator (in particular if we process a digital object).
Then a tracking algorithm (ii) extracts an approximate centerline by
following local maxima in the accumulation image. Finally, to remove
digitization effects due to the 3D discrete accumulation image, we
optimize the position of center points (iii) by a gradient descent
method.

\subsection{Accumulation Images From Normal Vectors}

The first algorithm requires as input a set of faces (a mesh or a
digital surface), their associated normal vectors, and a 3D digital
space (the 3D grid that will store the accumulated values). If the
input object is a mesh, the gridstep of the digitization grid must be
specified by the user. Depending on the awaited accuracy, a default
gridstep can be chosen as the median size of mesh faces (with the face
size defined as its longest edge). If the input object is a digital
object or a heightmap, the digitization grid just matches their
resolution. Besides, this algorithm requires as parameter some
approximation of the tube radius $R$.

\begin{figure}
  \centering
\begin{tabular}{cccc}
\psfrag{am}[c][][1]{\hskip 0.2cm \textbf{\textcolor{lblue}{$\alpha_{min}$}}}
\psfrag{R}[c][][1]{\hskip 0.2cm \textbf{\textcolor{gray}{$R$}}}
\psfrag{ep}[c][][1]{\hskip 0.2cm \textbf{\textcolor{gray}{$\epsilon$}}}
\psfrag{fj}[c][][1]{\hskip 0.2cm \textbf{\textcolor{dorange}{$f_j$}}}
\psfrag{nj}[c][][1]{\hskip 0.2cm \textbf{\textcolor{dorange}{$\overrightarrow{n_j}$}}}
\psfrag{fk}[c][][1]{\hskip 0.2cm \textbf{\textcolor{dgreen}{$f_k$}}}
\psfrag{nk}[c][][1]{\hskip 0.2cm \textbf{\textcolor{dgreen}{$\overrightarrow{n_k}$}}}
\includegraphics[height=0.22\textwidth]{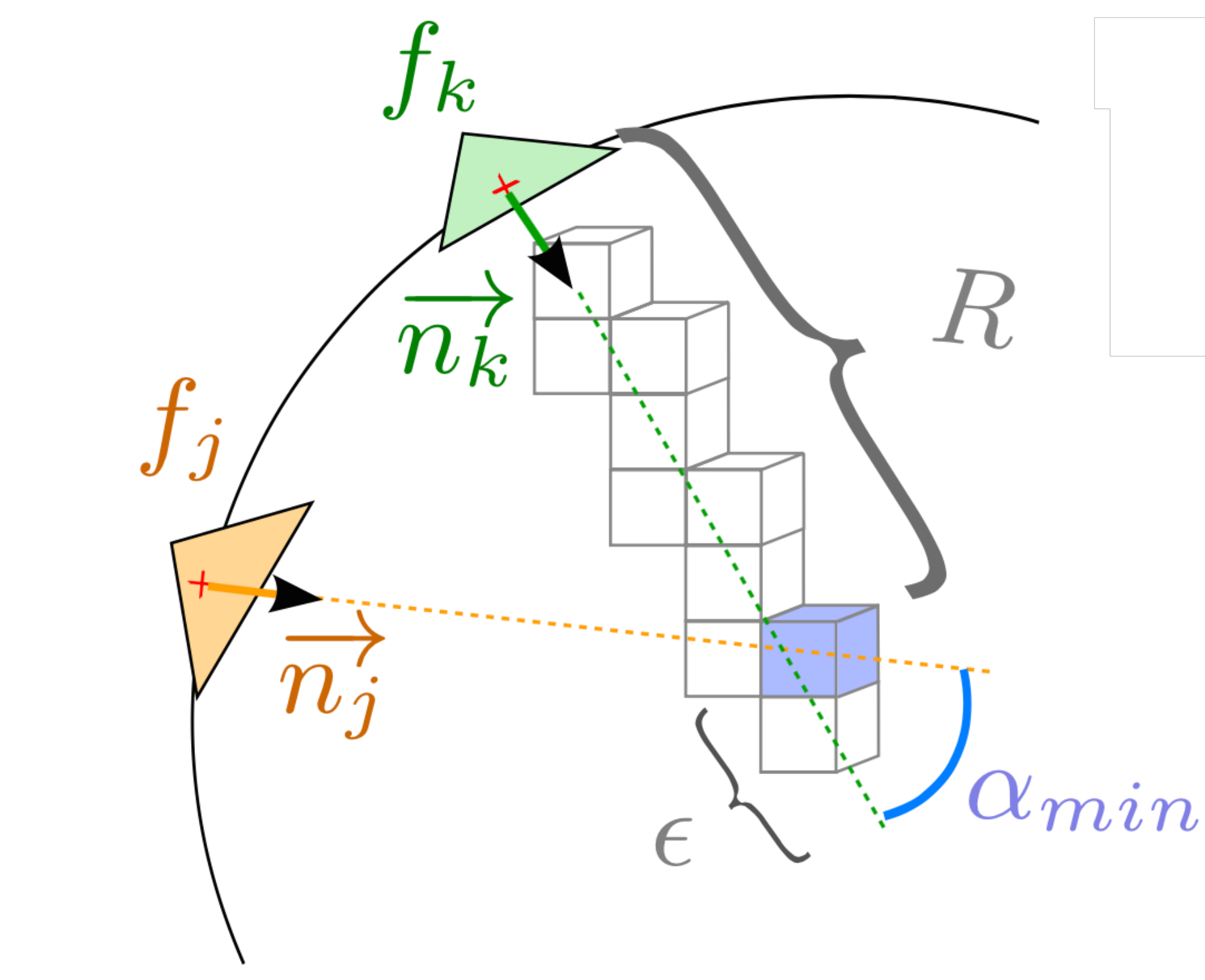}&
\includegraphics[height=0.22\textwidth]{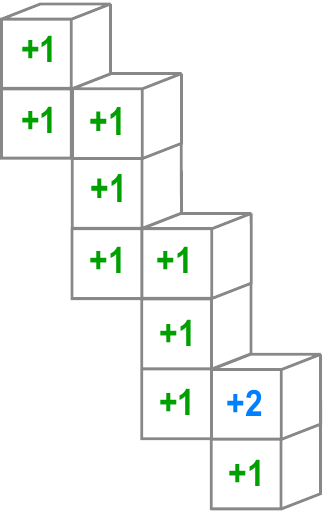}&
\psfrag{dk}[c][][1]{\hskip 0.2cm \textbf{\textcolor{lblue}{$\overrightarrow{d_k}$}}}
\psfrag{nj}[c][][1]{\hskip 0.2cm \textbf{\textcolor{dorange}{$\overrightarrow{n_j}$}}}
\psfrag{nk}[c][][1]{\hskip 0.2cm \textbf{\textcolor{dgreen}{$\overrightarrow{n_k}$}}}
\includegraphics[height=0.22\textwidth]{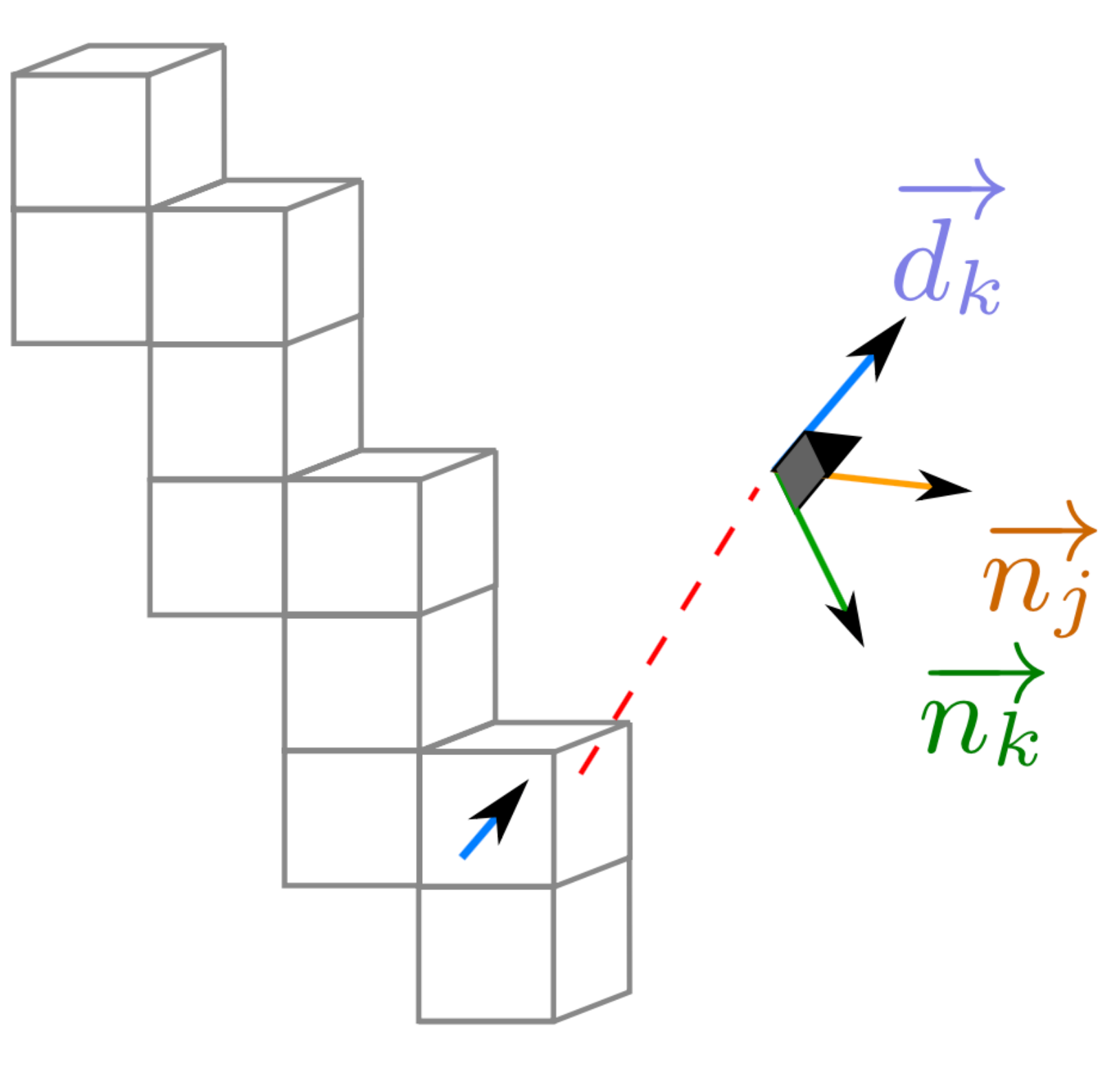}&

\includegraphics[height=0.22\textwidth]{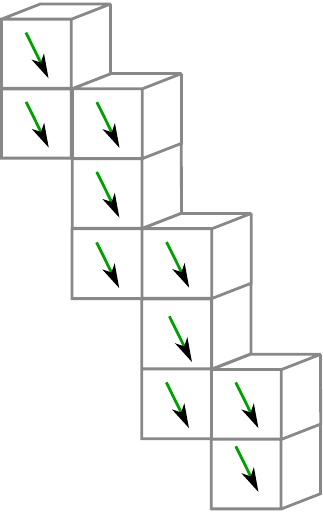}\\
{\scriptsize{(a) directional scan}}    & {\scriptsize{ (b) \texttt{accImage}}} &{ \scriptsize{ (c) \texttt{dirImage} }}&{\scriptsize{ (d) \texttt{lastVectors}}}
\end{tabular}

\caption{Illustration of \RefAlgo{AlgoImageAcc}, which builds an
  accumulation image whose peaks match the centerline of the tubular
  shape.  }
\label{fig:accumulation}

\end{figure}

The whole algorithm is detailed in \RefAlgo{AlgoImageAcc}. It outputs
for each voxel the number of normal vectors going through it as well
as a vector estimating the tube local main directions (i.e. the
tangent to the centerline or equivalently the direction of minimal
curvature along the tube boundary). \RefFigure{fig:accumulation}
illustrates the main steps of the algorithm with the 3D directional
scans, starting from the face origin $f_k$ in the direction of its
normal vector $\overrightarrow{n_k}$ along a distance denoted
\texttt{accRadius} (set to $R+\epsilon$ where $\epsilon$ is used to
take into account possible small variations of the radius along the
tube, see image (a) of \RefFigure{AlgoImageAcc}). During the scan, the
accumulation scores are stored for each visited voxel (image (b) of
the same figure). The principal direction $\overrightarrow{p}$ of a
voxel is also updated for each scan (image (c)). More precisely, if we
denote by $\overrightarrow{n_j}$ and $\overrightarrow{n_k}$ the two
last normal vectors intersecting a voxel $V$ for the scans $j$ and
$k$, the principal direction $\overrightarrow{d_k}$ for the current
scan is given by: $ \overrightarrow{d_k} = \overrightarrow{d_j} + (
\overrightarrow{n_{k}} \wedge \overrightarrow{n_{j}})$.  In order to
ignore non significant directions induced by near colinear vectors, we
add a small constant (set by default to $0.1$) to filter the norm of
the resulting vector $ \overrightarrow{n_{k}} \wedge
\overrightarrow{n_{j}}$.

\begin{figure}
\center
\begin{tabular}{cccc}
\includegraphics[width=0.24\textwidth]{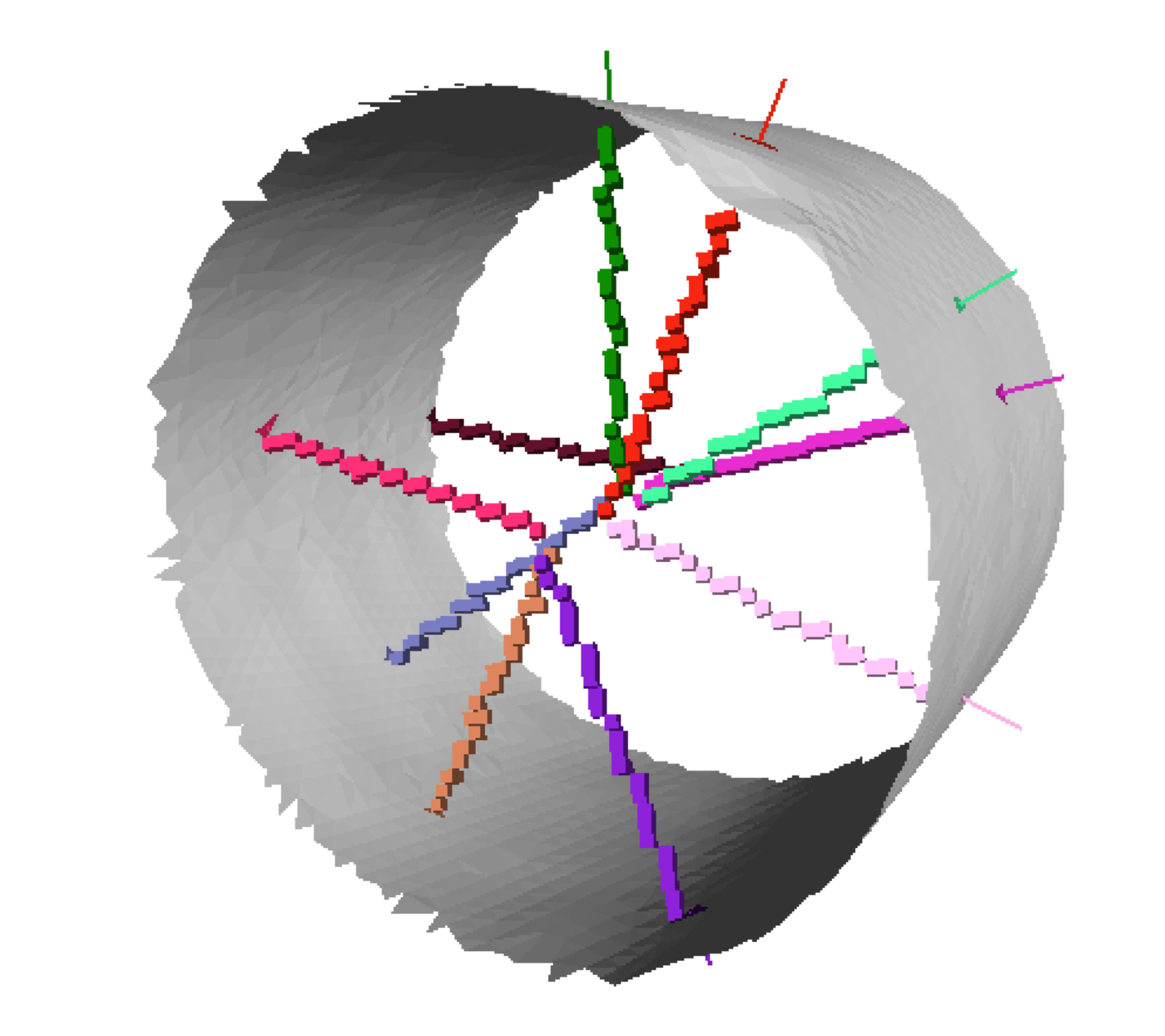}&
\includegraphics[width=0.24\textwidth]{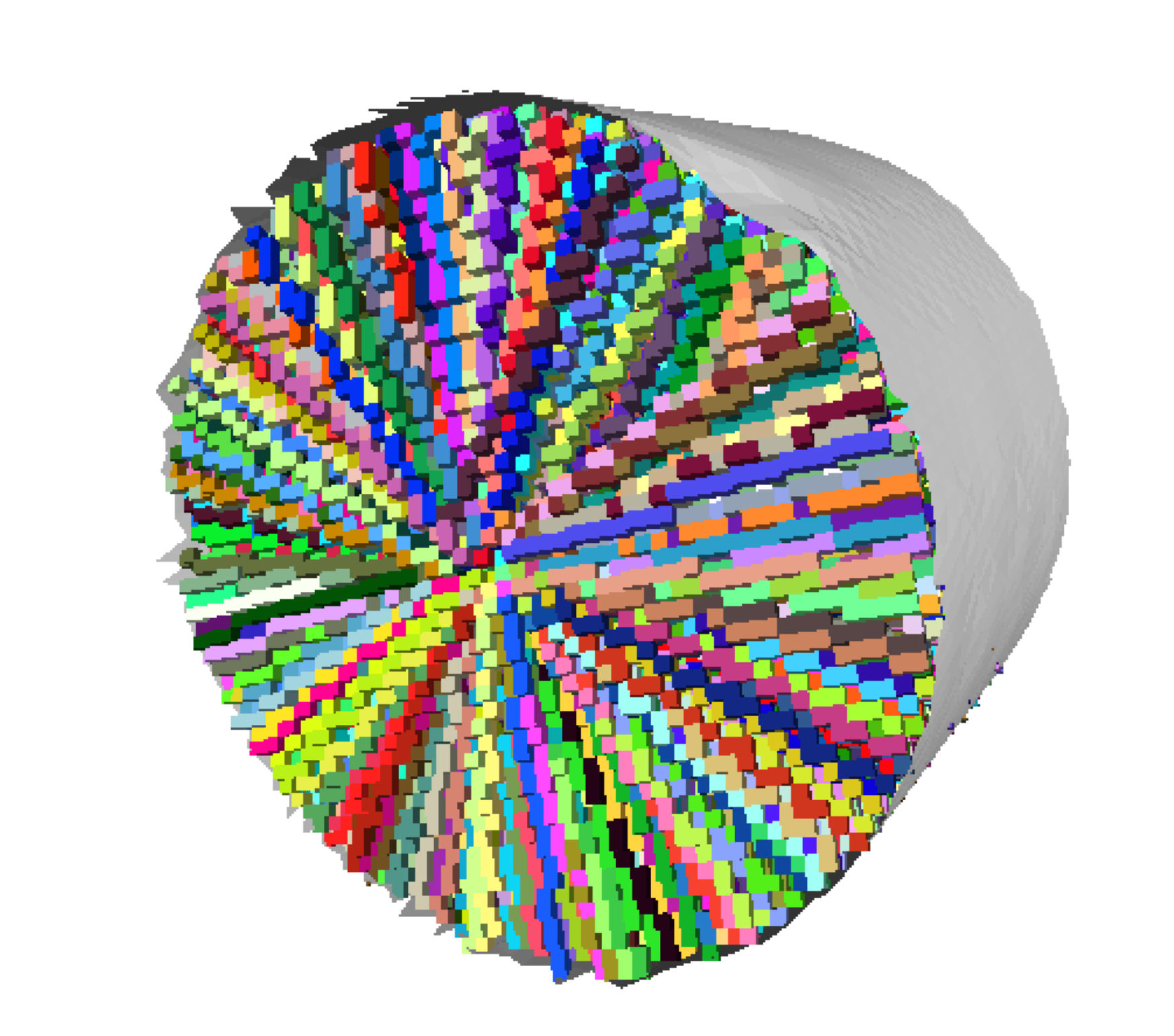}&
\includegraphics[width=0.24\textwidth]{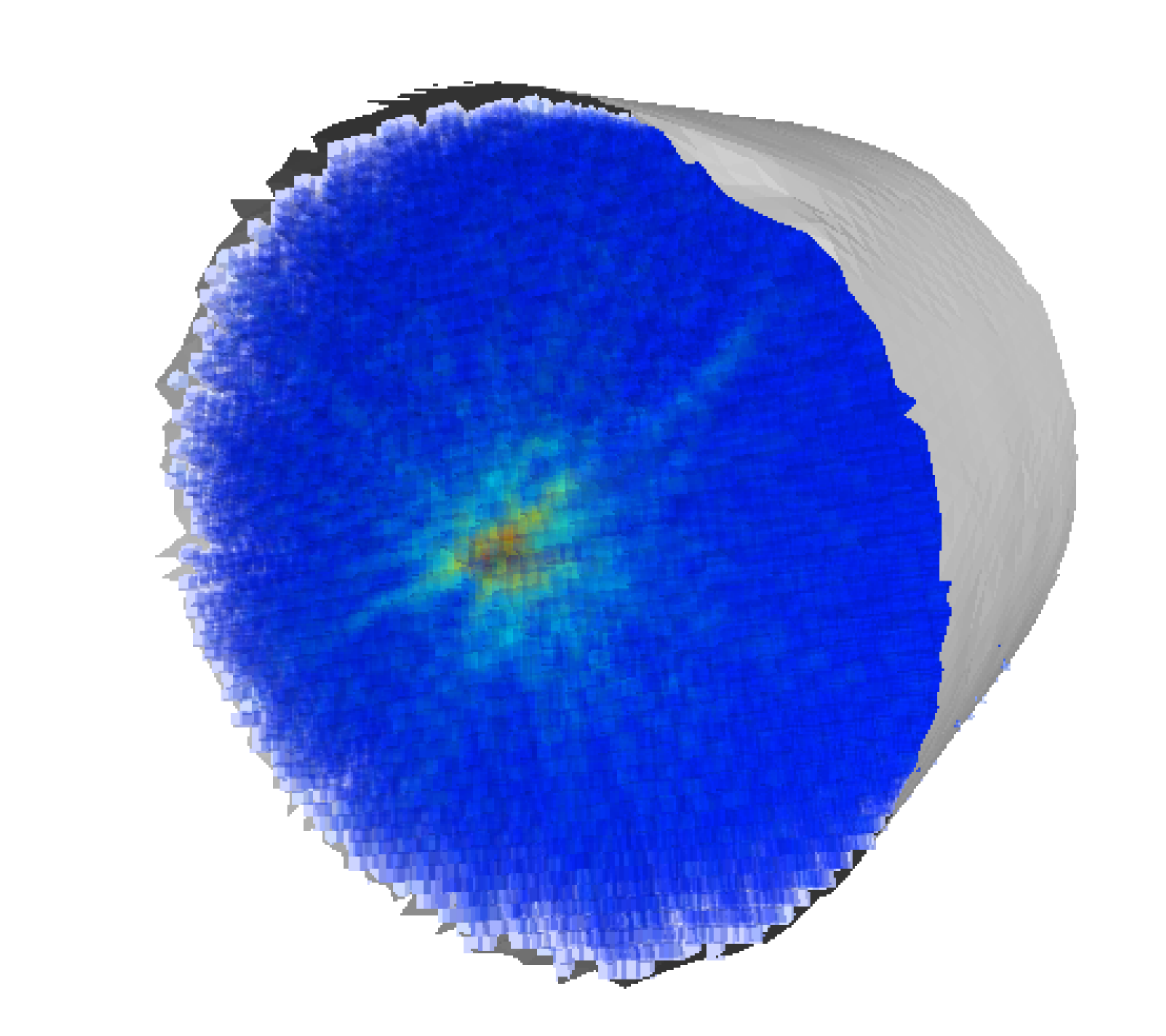}&
\includegraphics[width=0.24\textwidth]{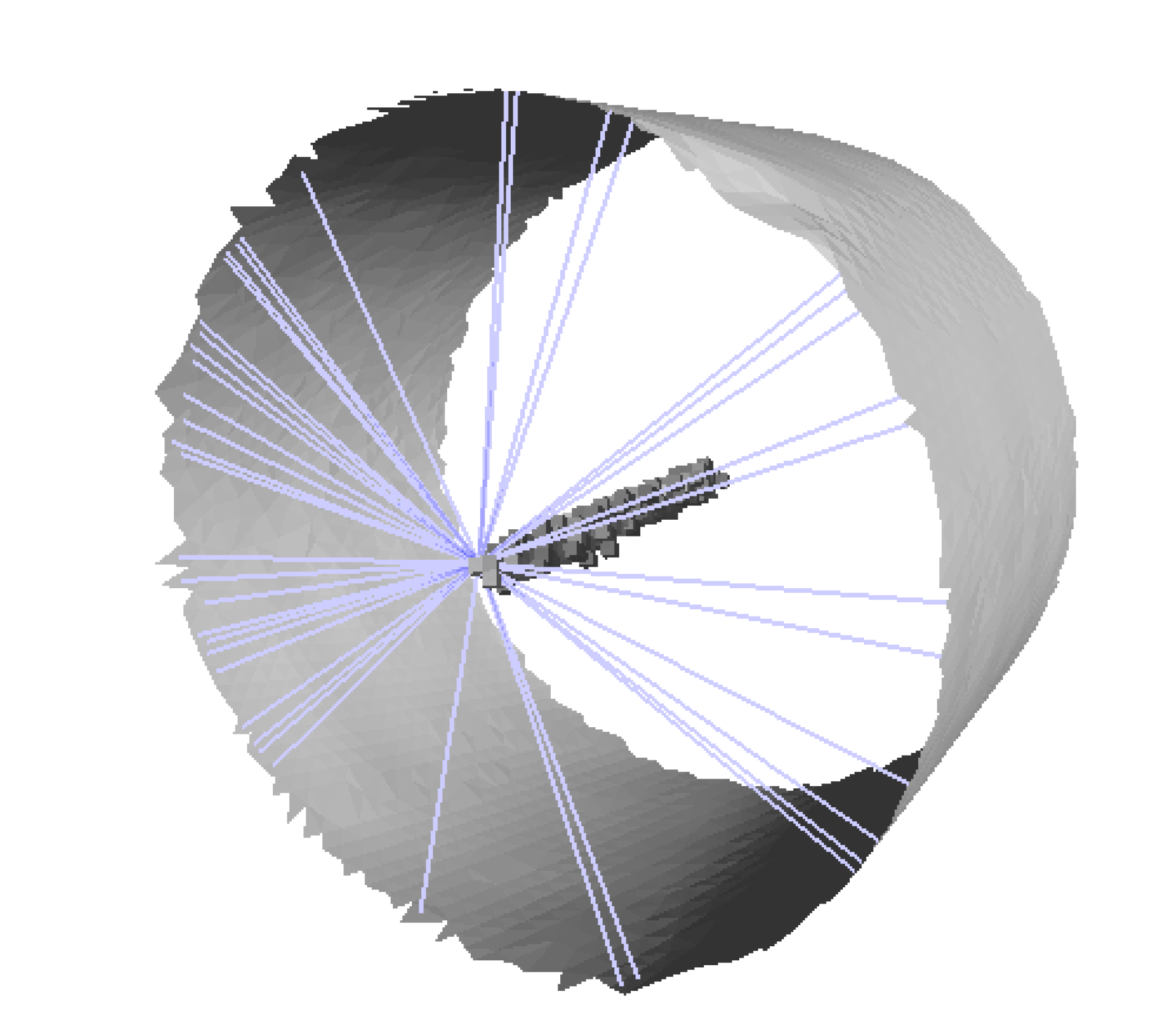}\\
(a) & (b) & (c) & (d)
\end{tabular}
\caption{Illustration of accumulation images generated from surface
  normal vectors. Image (a) (resp. (b)) illustrates some (resp. all)
  scanning directions defined from mesh triangles. Image (c)
  illustrates the values obtained in the 3D accumulation image and (d)
  shows some voxels having accumulation score upper than a given
  threshold. In the last image we also display the set of faces
  contributing to the accumulation score.}
\label{FigIllustrationAccumulation}
\end{figure}

\begin{figure}
\center
\begin{tabular}{ccccc}
\includegraphics[width=0.2\textwidth, trim = 180 100 100 30, clip=true]{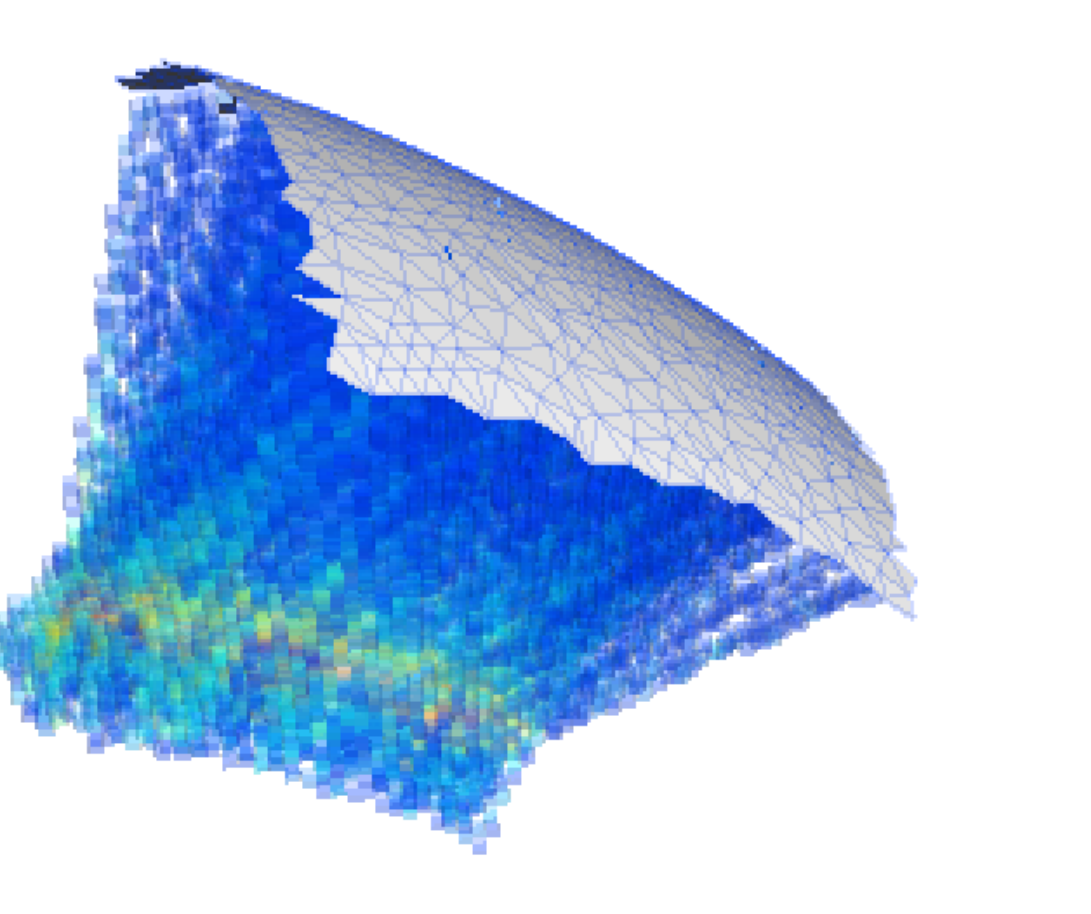}&
\includegraphics[width=0.2\textwidth]{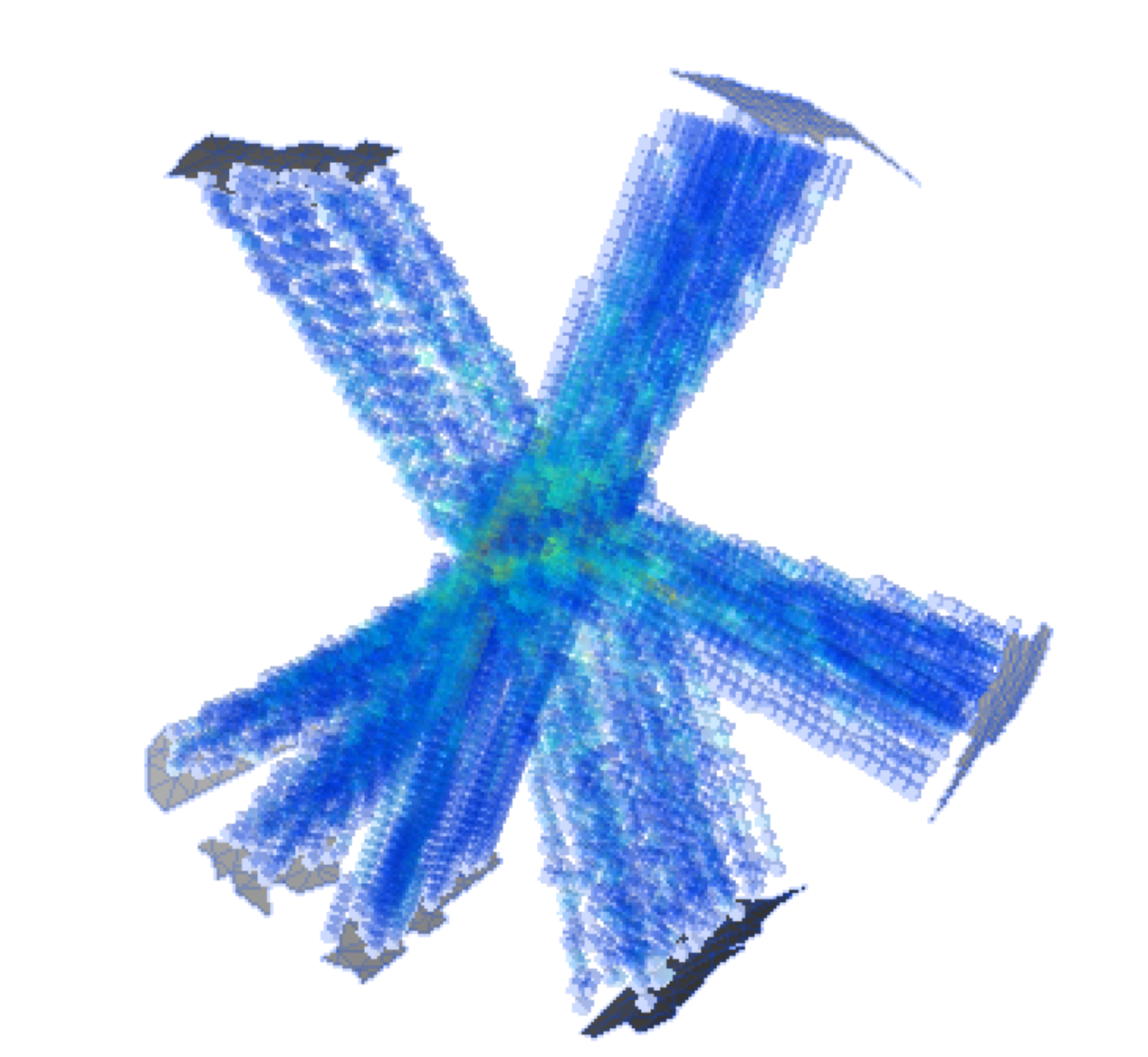}&
\includegraphics[width=0.2\textwidth, trim = 100  50 70 40, clip= true]{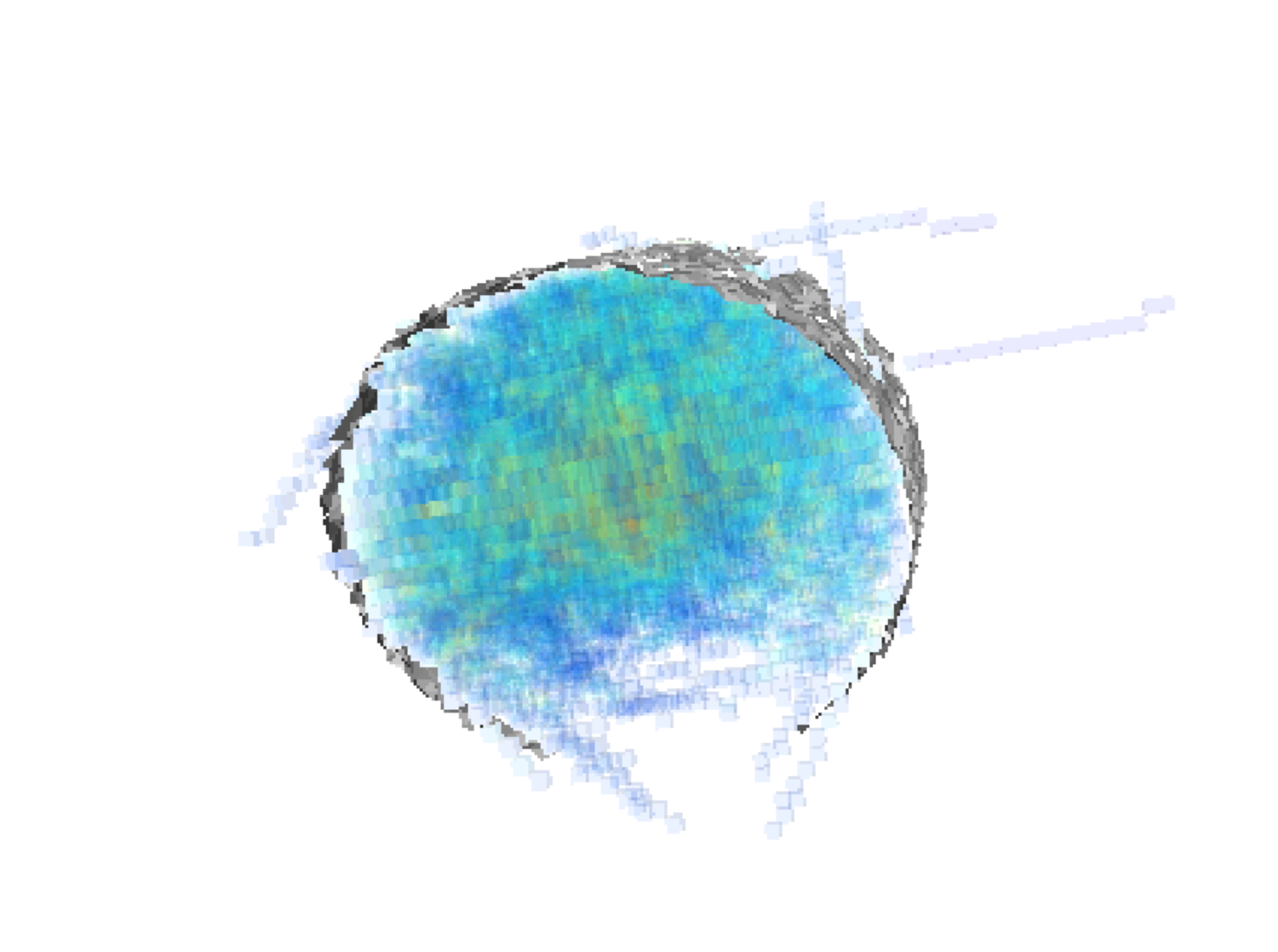}&
\includegraphics[width=0.2\textwidth]{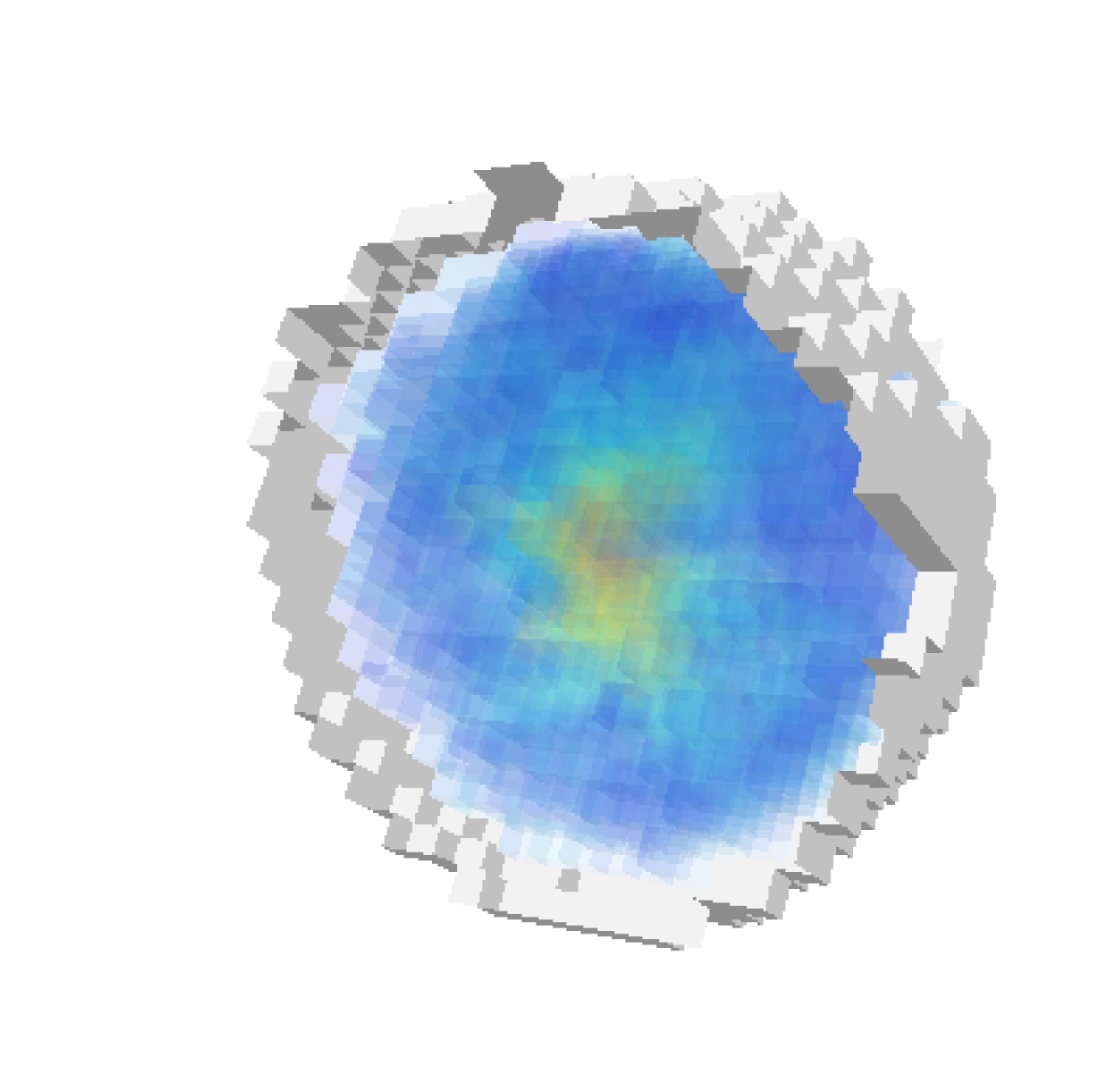}&
\includegraphics[width=0.2\textwidth]{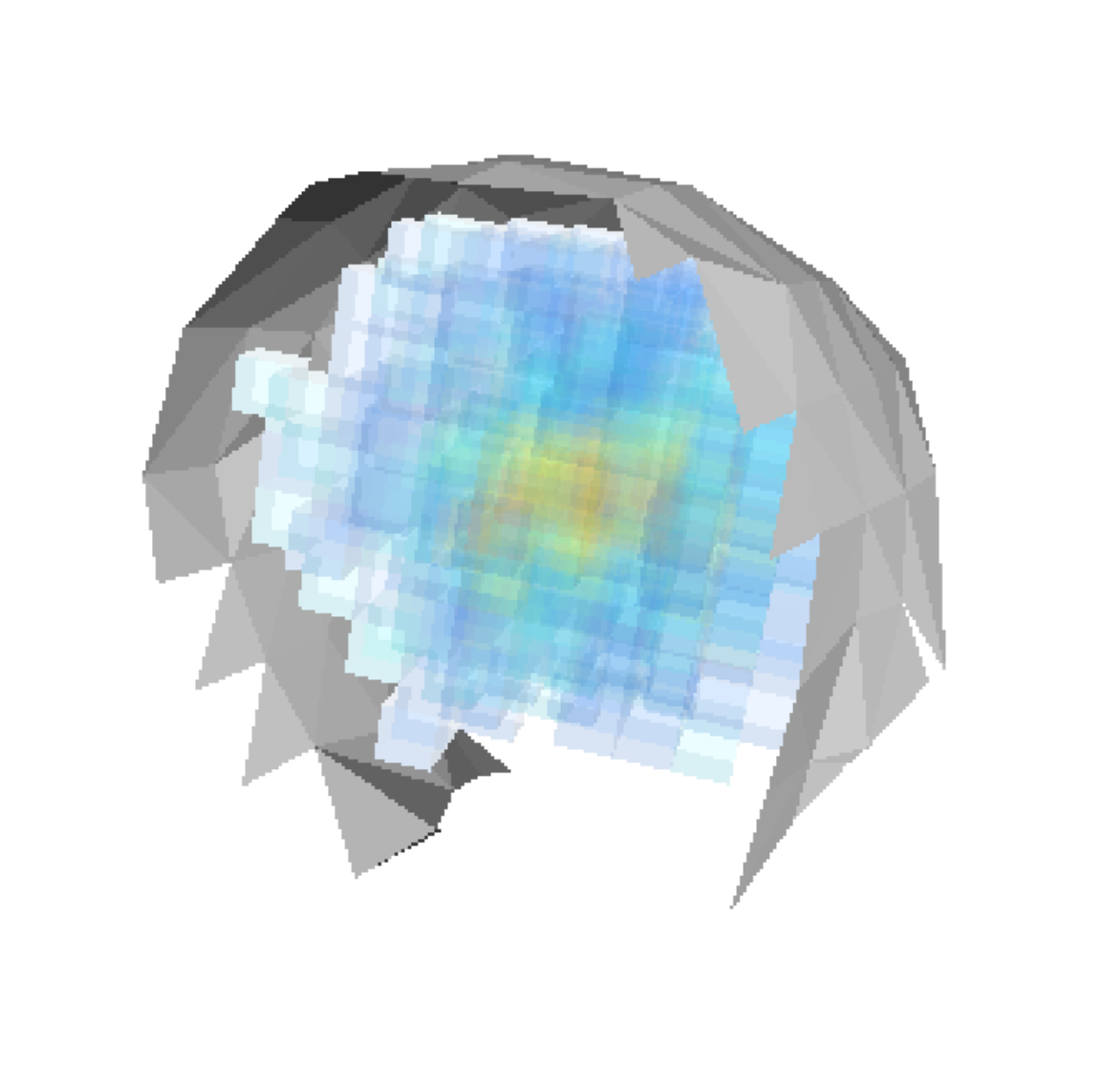}\\
(a) & (b) & (c) & (d) & (e) \\
\includegraphics[width=0.2\textwidth]{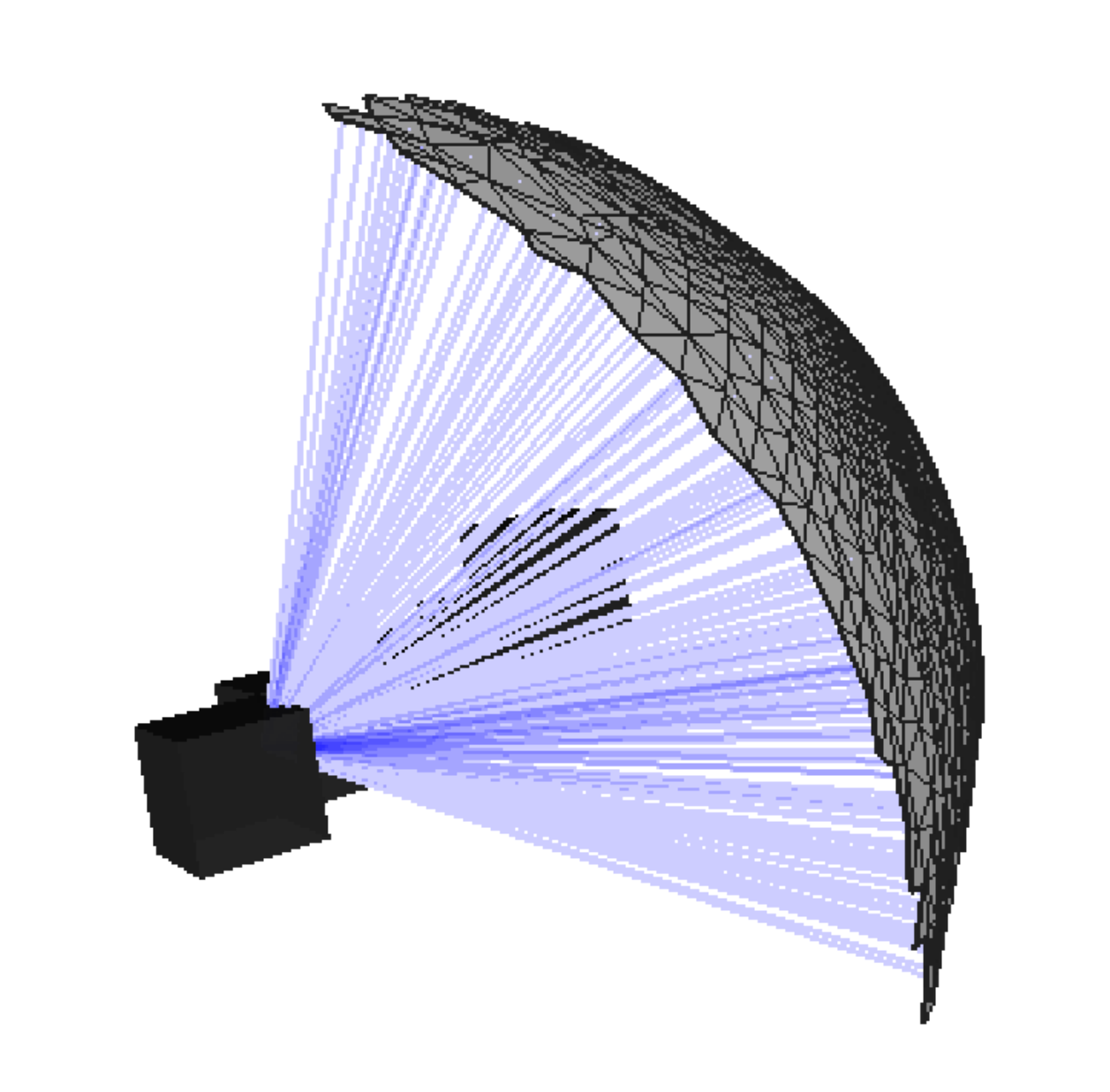}&
\includegraphics[width=0.2\textwidth]{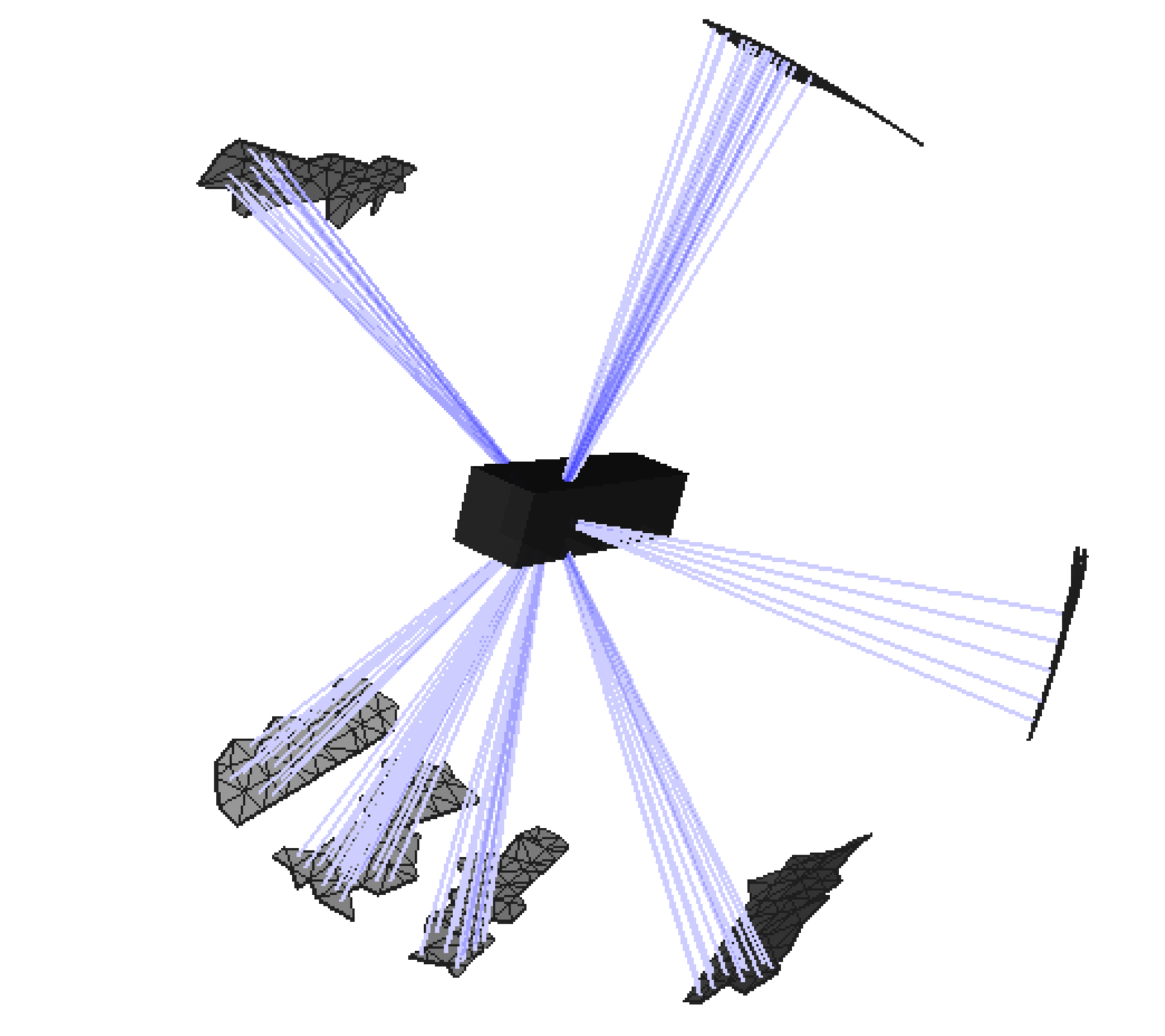}&
\includegraphics[width=0.2\textwidth]{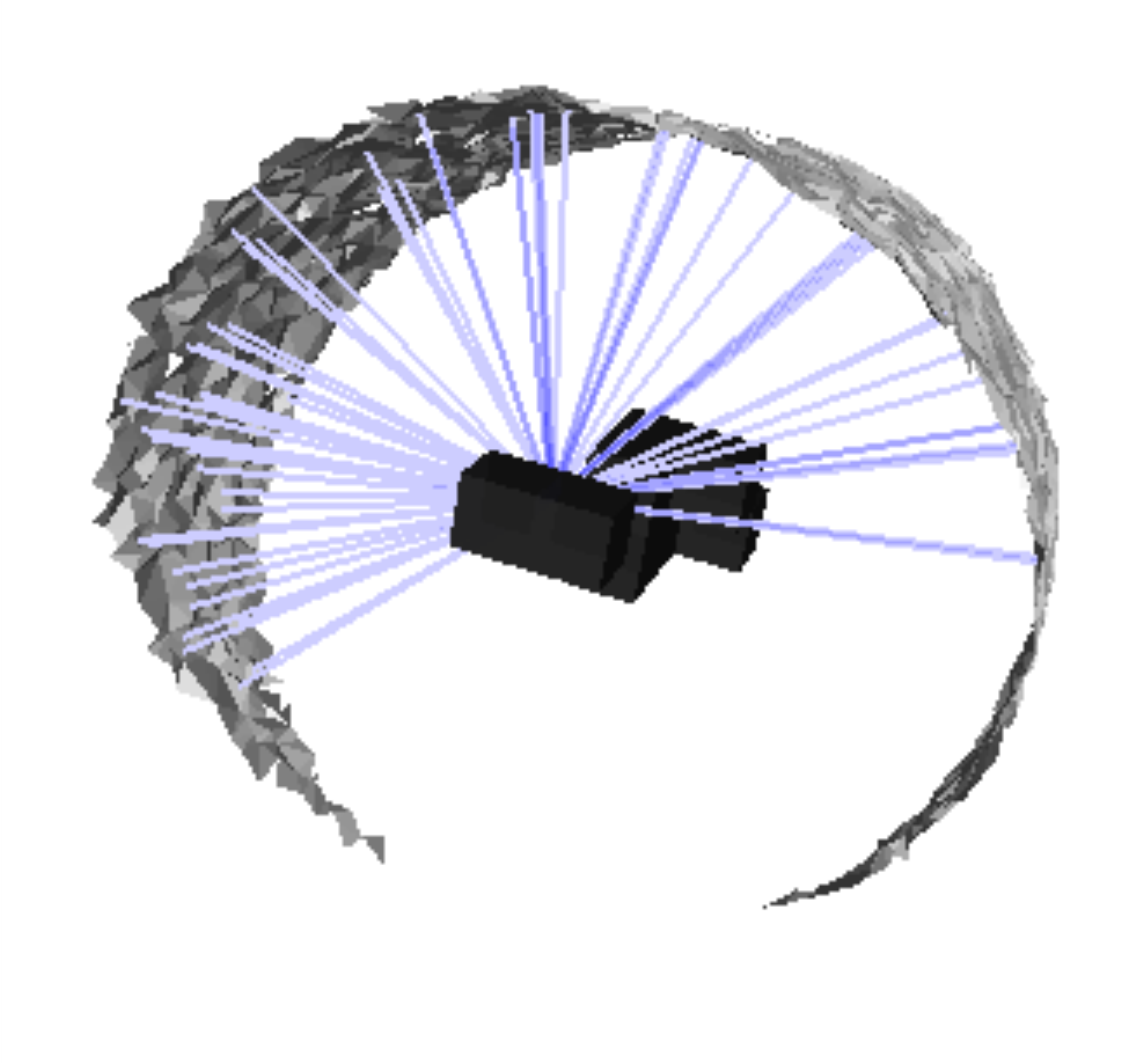}&
\includegraphics[width=0.2\textwidth]{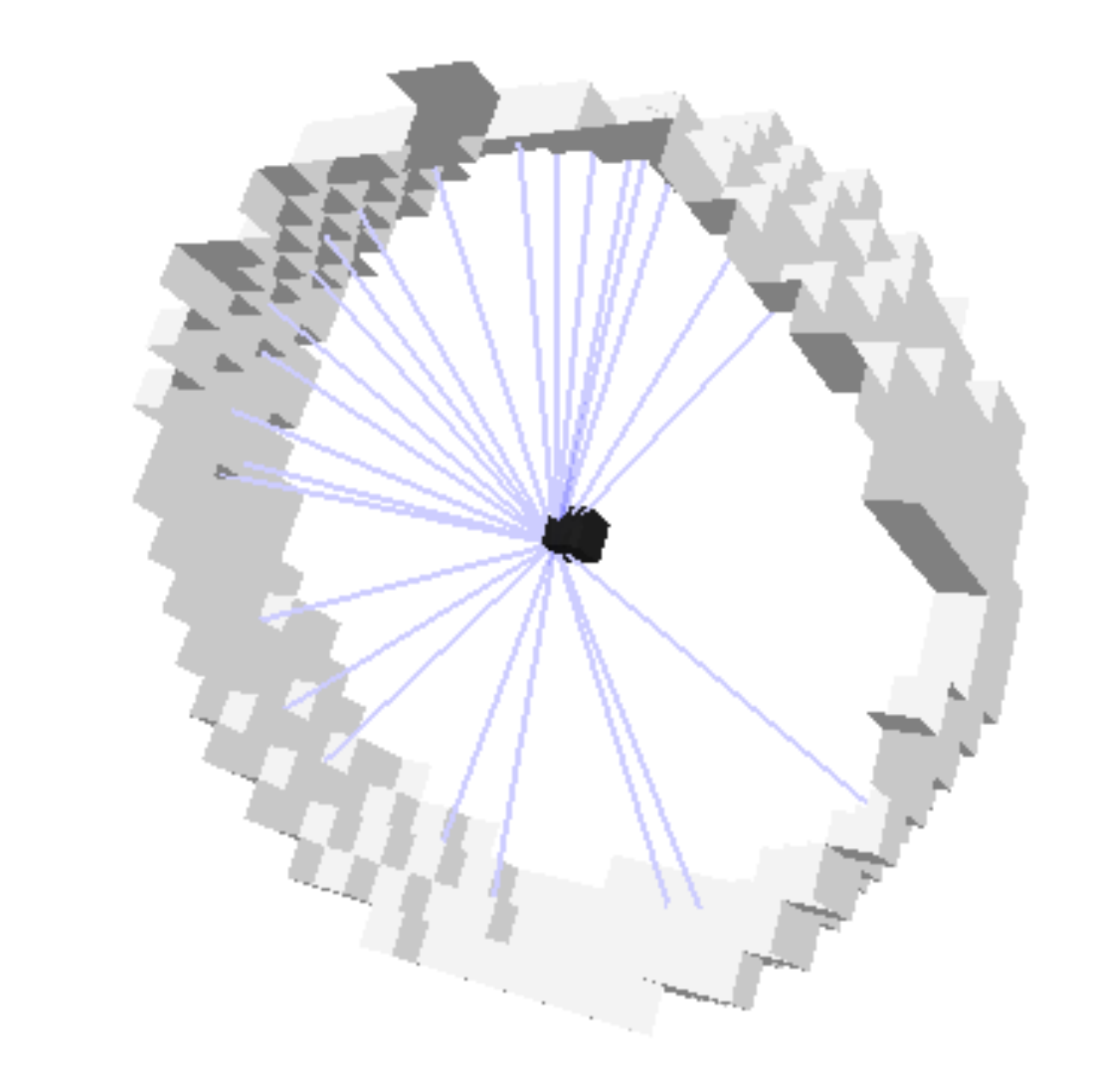}&
\includegraphics[width=0.2\textwidth]{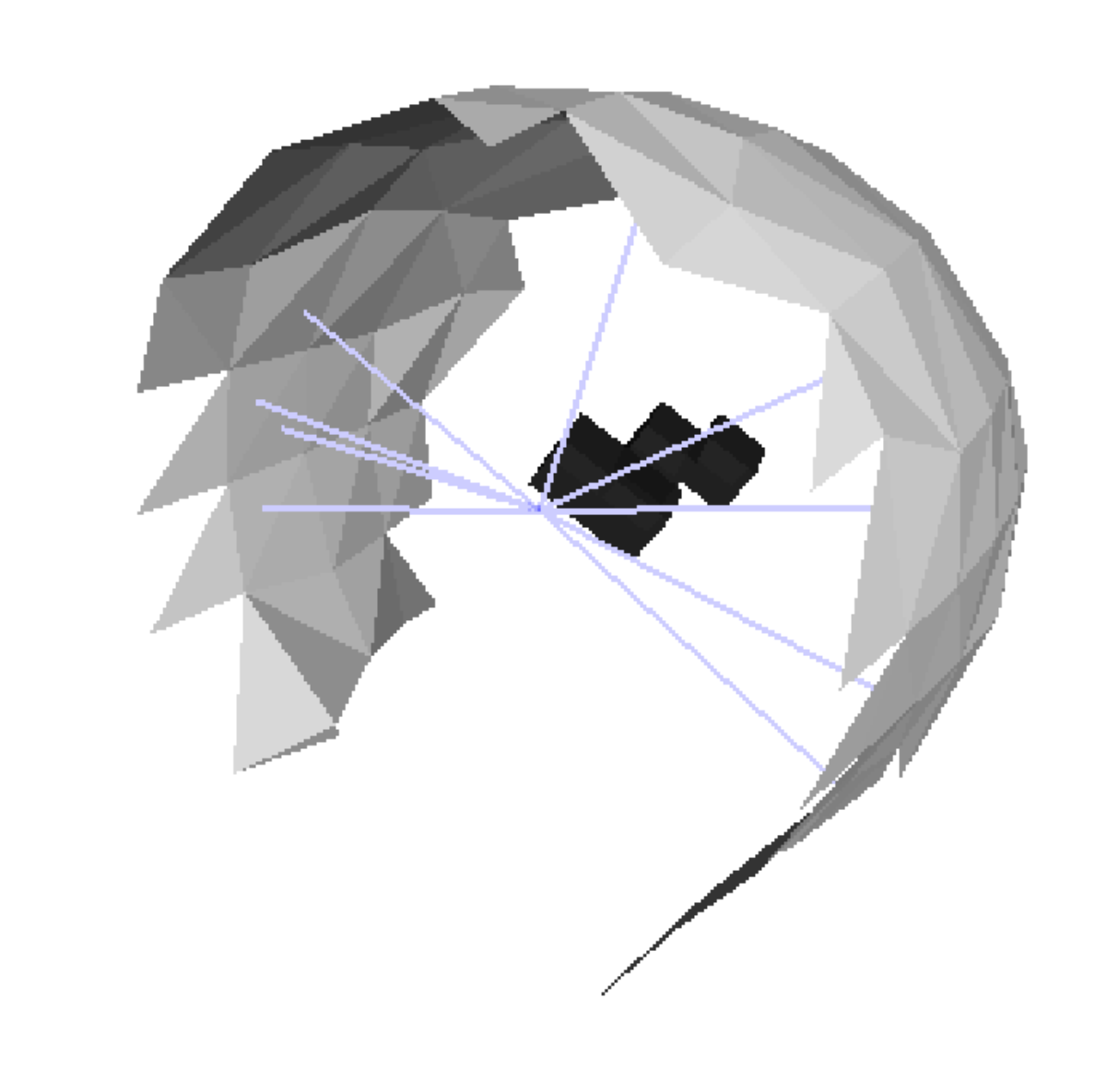}\\
(f) & (g) & (h) & (i) & (j) \\
\end{tabular}
\caption{Experiment of the robustness of the surface normal
  accumulation algorithm applied on different types of input surface: on
  sector filtered mesh (a,f), on partial scan mesh (b,g), on
  noisy mesh (c,h), on digital surface (d,i), and on small resolution
  mesh (e,j). }
\label{FigExpeRobustessAcc}
\end{figure}

\RefFigure{FigIllustrationAccumulation} illustrates the computations
made in \RefAlgo{AlgoImageAcc} for a mesh input data, and it shows the
resulting accumulation image (image (c)) and vectors
$\overrightarrow{n}$ orthogonal to the tube main direction
$\overrightarrow{d}$. Since centerline extraction relies on these
accumulation images, we evaluate the robustness of these images with
various input surface types. The first row of
\RefFigure{FigExpeRobustessAcc} presents the resulting 3D accumulation
images obtained on partial, noisy, digital or on small resolution
mesh. In all these configurations, relative maximal values are indeed
well located near the center of the tubular shape. A fixed threshold
was applied in order to highlight the voxels with accumulation values
close to maximal ones. Such voxels are drawn in black and for a
particular selected voxel, we have highlighted their scanning origin
faces with blue lines. All these results confirm the robustness of the
proposed algorithm. We have therefore a solid basis for the tracking
algorithm presented in the following part.

\begin{algorithm}
\begin{scriptsize}
\SetKwInOut{Input}{Input} \SetKwInOut{Output}{Ouput} \SetKwInOut{Variable}{Variable}
\Input{\texttt{mesh} // Triangular mesh of a tube \\
       \texttt{accRadius} // Accumulation length from center of faces \\
       \texttt{minNorm = 0.1} // Minimum norm value
}
\Output{\texttt{accImage} // Accumulation of normal vector number passing through a coordinate \\
       \texttt{dirImage} // Cross product of all normals passing through a coordinate \\
        \texttt{maxAcc} // Maximum number of normals passing through an (x,y,z) coordinate \\
        \texttt{maxPt} // maxAcc coordinates
}
\Variable{\texttt{lastVectors} // The last considered normal for each (x,y,z) coordinate\\
          \texttt{mainAxis} // Vector contributing to the cross product of a directional vector
}

\texttt{lastVectors = Image3D(mesh.dimensions())}

\texttt{maxAcc = 0}

\ForEach{\texttt{face} in \texttt{mesh}}
    {
      \texttt{currentPt = face.center}

      \texttt{normalVector = face.normalVector().normalized()}

      \While{\texttt{distance(currentPt, face.center) $<$ accRadius}}
      {

        \If {\texttt{accImage[currentPt] != 0} }
        {

          \texttt{mainAxis = lastVectors[currentPt] $\times$ normalVector}

          \If {\texttt{norm(mainAxis) $>$ minNorm}}
              {
                \texttt{dirImage[currentPt] += mainAxis*sign(mainAxis $\bullet$ dirImage[currentPt])}
              }
        }
        \texttt{lastVectors[currentPt] = normalVector}

        \texttt{accImage[currentPt]++}

        \If {\texttt{accImage[currentPt] $>$ maxAcc}}
            {

              \texttt{maxAcc = accImage[currentPt]}

              \texttt{maxPt = currentPt}

            }

        \texttt{currentPt += normalVector}
      }
    }
\end{scriptsize}
\caption{\texttt{accumulationFromNormalVectors} : From position and normals of faces of an input mesh, this
  algorithm computes an accumulation image (\texttt{accImage}) by a
  directional scan starting from a face center in the direction of the
  face inward normal. It also outputs the image of vectors
  representing the local main axis direction of the tubular shape
  (\texttt{dirImage}).}

\label{AlgoImageAcc}

\end{algorithm}

\subsection{Centerline Tracking from Image Accumulation }

\begin{floatingfigure}[r]{4cm}
\vskip -0.3cm
\hskip -0.3cm
\begin{center}
\psfrag{25}[c][][0.6]{\hskip 0.2cm \textbf{\textcolor{blue}{$\sigma=25$}}}
\psfrag{50}[c][][0.6]{\hskip 0.2cm \textbf{\textcolor{blue}{$\sigma=50$}}}
\psfrag{100}[c][][0.6]{\hskip 0.2cm \textbf{\textcolor{blue}{$\sigma=100$}}}
\psfrag{150}[c][][0.6]{\hskip 0.2cm \textbf{\textcolor{blue}{$\sigma=150$}}}
\includegraphics[width=0.3\textwidth]{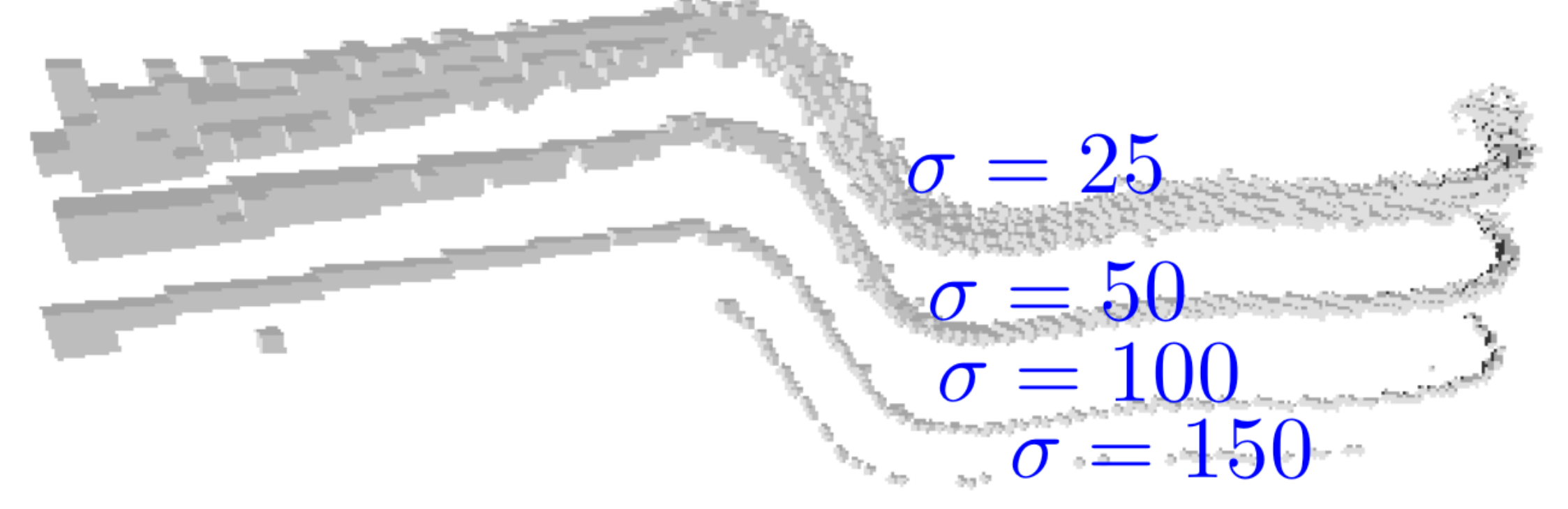}\\
\end{center}
\end{floatingfigure}
Even if the maximal values obtained in the previous part are well
centered on the tubular object, a simple thresholding is not robust
enough to extract directly the centerline. Furthermore it implies the
manual adjustment of the threshold parameter. To illustrate this
point, the image on the side shows different results obtained by
choosing various threshold parameters $\sigma$. A too strict threshold
implies disconnected points, while a less restrictive one produces a
thick line with parasite voxels.

\begin{figure}
\centering
\psfrag{Ci}[c][][1]{\textbf{\textcolor{yellow}{$C_i$}}}
\psfrag{Ci+1}[c][][1]{\textbf{\textcolor{yellow}{$C_{i+1}$}}}
\psfrag{trac}[c][][.7]{\hspace{3em}\texttt{\textcolor{black}{trackStep}}}\psfrag{kStep}[c][][1]{}
\psfrag{dk}[c][][1]{\textbf{\textcolor{lblue}{$\overrightarrow{d_k}$}}}
\psfrag{dk+1}[c][][1]{\textbf{\textcolor{lblue}{$\overrightarrow{d_{k+1}}$}}}
\psfrag{Ipi}[c][][1]{$I_{patch}^{i}$}
\psfrag{Ipi+1}[c][][1]{$I_{patch}^{i+1}$}
\includegraphics[height=0.3\textwidth]{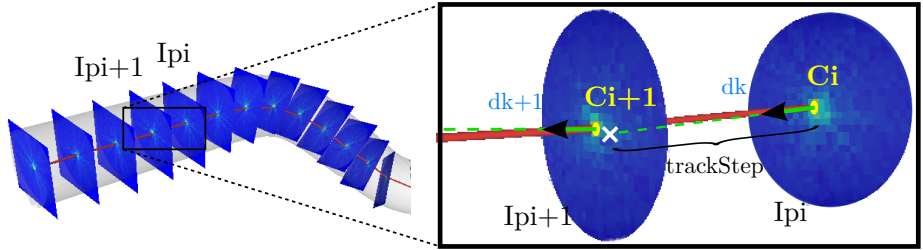}
\caption{Tracking algorithm step. The patch $I_{patch}^{i+1}$ is generated from the maximum $C_i$ of $I_{patch}^{i}$ in the $\protect\overrightarrow{d_k}$ direction at a \texttt{trackStep} distance before to localize the maximum $C_{i+1}$ of this new patch.}
\label{FigTracking3D}
\end{figure}

To better approach the centerline we propose to define a simple
tracking algorithm exploiting the output of \RefAlgo{AlgoImageAcc},
i.e. the accumulation image and the direction vectors image.  As
described in \RefAlgo{AlgoPatchTrackDir}, the main idea is to start
from a point $C_0$ detected as a maximal accumulation value of the 3D
accumulation image. Then, from a current point $C_i$ of the centerline,
the algorithm determines next point $C_{i+1}$ as the point
having maximal accumulation value in the 2D patch image $I_{patch}^i$
defined in the plane normal to the direction \texttt{dirImage}($C_i$)
at distance \texttt{trackStep} (see \RefFigure{FigTracking3D} (a,b)).

\begin{algorithm}
\begin{scriptsize}
\SetKwInOut{Input}{Input} \SetKwInOut{Output}{Output} \SetKwInOut{Variable}{Variable}

\Input{\texttt{accImage} // Accumulation of normal vector number passing through a coordinate \\
       \texttt{dirImage} // Cross product of all normals passing through a coordinate \\
       \texttt{accRadius} // Accumulation length from center of faces \\
       \texttt{startPt}, // Start point for tracking (must belong to the centerline) \\
       \texttt{trackInFront} // True if tracking direction is in the startPt vector direction \\
       \texttt{trackStep} // Distance between two consecutive centerline points
}
\Output{\texttt{centerline} // Point set constituting the tube centerline }

\Variable{\texttt{continueTracking} // True if tracking can continue\\
          \texttt{patchSize} // Dimension of the square patch\\
          \texttt{currentPt, previousPt} // Considered point during an iteration \\
          \texttt{lastVect} // Directional vector associated to previousPt\\
          \texttt{centerPatch} // Patch center finding from currentPt \\
}

\texttt{patchImageSize} = 2 * aRadius \;

\texttt{centerline = emptySet()}

\texttt{continueTracking = true}

\texttt{patchSize = 2 * accRadius}

\texttt{currentPt = startPt}

\texttt{lastVect =  trackInFront ? dirImage( startPt ) : - dirImage( startPt ) }

\texttt{previousPt = startPt - lastVect * trackStep}

\While{\texttt{continueTracking}}
{
  \texttt{centerline.append( currentPt )}

  \texttt{dirVect = dirImage[currentPt].normalized()}

  \If{\texttt{lastVect.dot(dirVect) $< 0$}}
    {
      \texttt{dirVect = -dirVect}
    }

  \texttt{continueTracking = isInsideTube( accImage, currentPt, previousPt, trackStep, $\pi/3$ )}

  \texttt{previousPt = currentPt}

  // Defined the next image patch center point

  \texttt{centerPatch = currentPt + ( dirVect * trackStep )}

  \If {\texttt{not accImage.domain().contains( centerPatch )}}
    {
      break
    }

  // Extract a 2D image of size  2 * \texttt{accRadius} from the 3D image \texttt{accImage}, centered on \\
  // \texttt{centerPatch} and directed along \texttt{dirVect}\\
  \texttt{patchImage = extractPatch( accImage, centerPatch, dirVect, 2 * accRadius )}

  \texttt{maxCoords = getMaxCoords( patchImage )}

  \texttt{lastVect = dirVect}

  \texttt{previousPt = currentPt}

  \texttt{currentPt = patchSpaceToAccImageSpace( maxCoords)}

}
\texttt{return centerline}

\caption{\textbf{trackPatchCenter}: tracking algorithm in one
  direction, given a starting point and an orientation.}.
\label{AlgoPatchTrackDir}
\end{scriptsize}

\end{algorithm}

\subsection{Skeleton position optimization}
\label{skeletonSection}

Since the resulting tracking skeleton is embedded in a digital space,
it suffers from digitization artefacts and is not perfectly centered
within the input mesh. Moreover, depending on normal mesh quality, the
tracking algorithm can potentially be influenced by perturbated normal
directions, and may deviate from the expected centerline. Such
perturbations can dramatically degrade the quality of upcoming
geometric analysis, and hence impose some unwanted post processing
tasks. To avoid such a difficulty, we propose to apply an optimization
algorithm in order to obtain a perfectly centered spine line.

\begin{figure}
\center
\begin{tabular}{cc}
\psfrag{CC}[c][][1]{\hskip 0.2cm \textbf{$C'$}}
\psfrag{C}[c][][1]{\hskip 0.2cm \textcolor{dred}{\textbf{$C$}}}
\psfrag{M0}[c][][1]{\hskip 0.2cm \textbf{$M_0$}}
\psfrag{M1}[c][][1]{\hskip 0.2cm \textbf{$M_1$}}
\psfrag{M2}[c][][1]{\hskip 0.2cm \textbf{$M_2$}}
\psfrag{M3}[c][][1]{\hskip 0.2cm \textbf{$M_3$}}
\psfrag{M4}[c][][1]{\hskip 0.2cm \textbf{$M_4$}}
\psfrag{P0}[c][][1]{\hskip 0.2cm \textcolor{dred}{\textbf{$P_0$}}}
\psfrag{P1}[c][][1]{\hskip 0.2cm \textcolor{dred}{\textbf{$P_1$}}}
\psfrag{P2}[c][][1]{\hskip 0.2cm \textcolor{dred}{\textbf{$P_2$}}}
\psfrag{P3}[c][][1]{\hskip 0.2cm \textcolor{dred}{\textbf{$P_3$}}}
\psfrag{P4}[c][][1]{\hskip 0.2cm \textcolor{dred}{\textbf{$P_4$}}}
\psfrag{f}[c][][1]{\hskip 0.2cm \textcolor{dred}{\textbf{$\overrightarrow{f}$}}}
\includegraphics[width=0.28\textwidth]{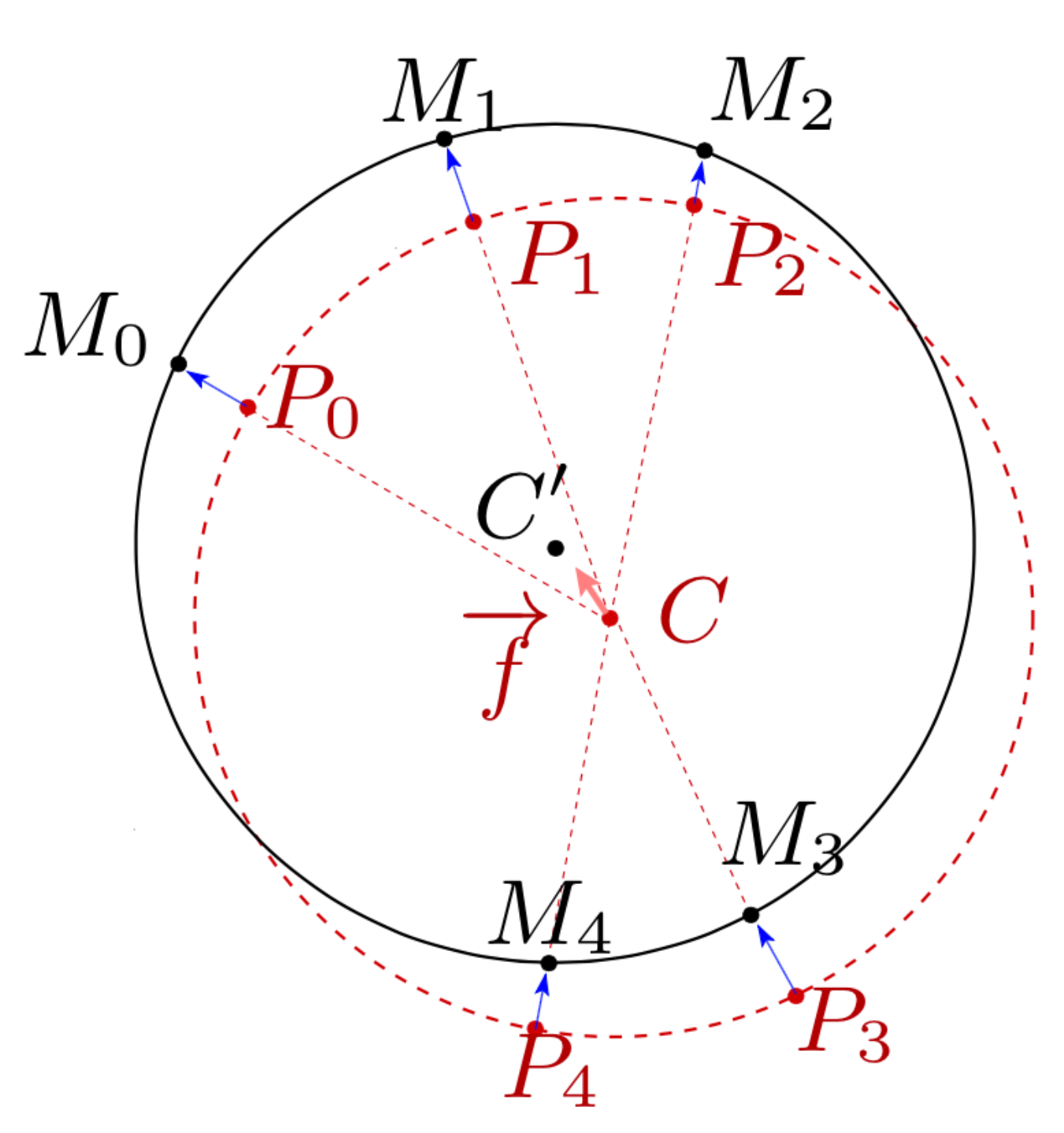}&
\includegraphics[width=0.4\textwidth]{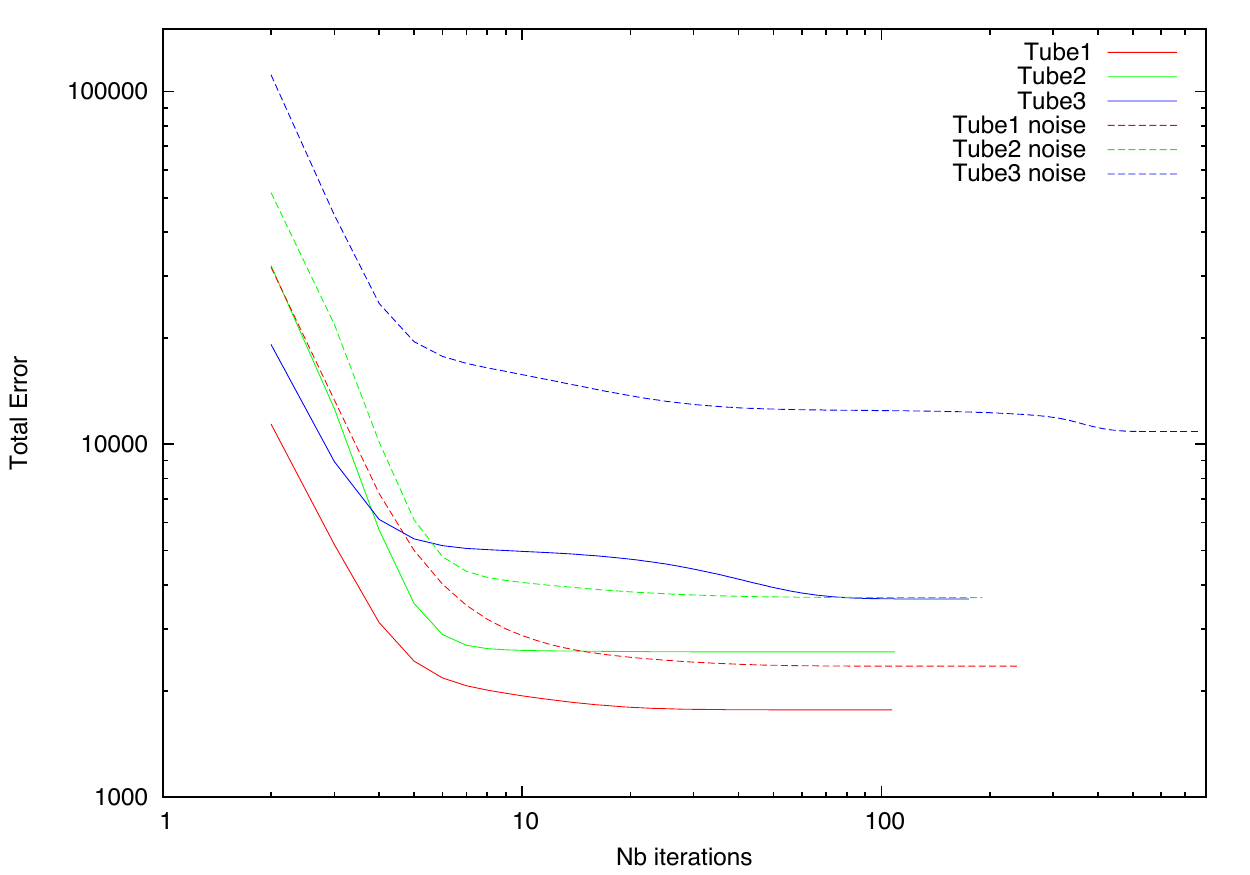}
\end{tabular}
\vskip -0.2cm
\caption{(left) Illustration of the process to optimize the centerline
  position (point $C$) with elastic forces (blue arrows). Each elastic
  force is attached to one point of the input mesh sector (a point
  $M_i$ represented in black) and oriented in the direction of the
  center of the virtual circle of center $S$. (right) evolution of the
  convergence speed in the optimization process.}
\label{FigForceCircle}
\end{figure}

The idea is to model the quality of the current fitting by an error
$E_s(C)$, defined as the sum of the squared difference between the
known tube radius $R$ and the distance between the tube center $C$ and
its associated input mesh points $M_i$. We wish to find the best
position for center $C$ that minimizes this error. Otherwise said, we
look for the circle of radius $R$ that best fits the data points $M_i$
in the least-square sense. Hence, the error is

\begin{equation}
E_s(C) = \sum_{i=0}^{N-1}{(\|\overrightarrow{CM_i}\|-R)^2}.
\end{equation}
This minimization problem is easily solved by a gradient descent
algorithm that follows the direction of steepest descent of the
error. By simple derivation, its gradient is
\begin{equation}
\nabla E_s(C) = 2\sum_{i=0}^{N-1} \frac{\overrightarrow{CM_i}}{\|\overrightarrow{CM_i}\|}{(R-\|\overrightarrow{CM_i}\|)}.
\end{equation}
The gradient descent can also be interpreted as elastic forces acting
on the center $C$ and pulling or pushing it in the direction of data
according to the current distance. Then the minimization process
applies at each step of the process the sum $ \overrightarrow{f}$ of
theses forces on $C$, giving with the notations of
Fig.~\ref{FigForceCircle}:   $\overrightarrow{f} = \sum_{i=0}^{N}{\overrightarrow{P_iM_i}}$

By this way, at each step, the total error $E_S$ decrease and we
iterate the process until convergence, i.e. the difference of errors
between two iterations is below a fixed $\epsilon_o$.

Contrary to a simple average of neighborhood points, the optimization
performs well even on partial mesh data, with missing parts or
holes. Moreover, it is possible to ponder each force with its face
area in order to better balance forces in presence of irregular
sampling with variable density.

\section{Reconstruction Results and Geometric Analysis}

\begin{figure}
\centering
\begin{tabular}{cc}

\includegraphics[width=0.401\textwidth]{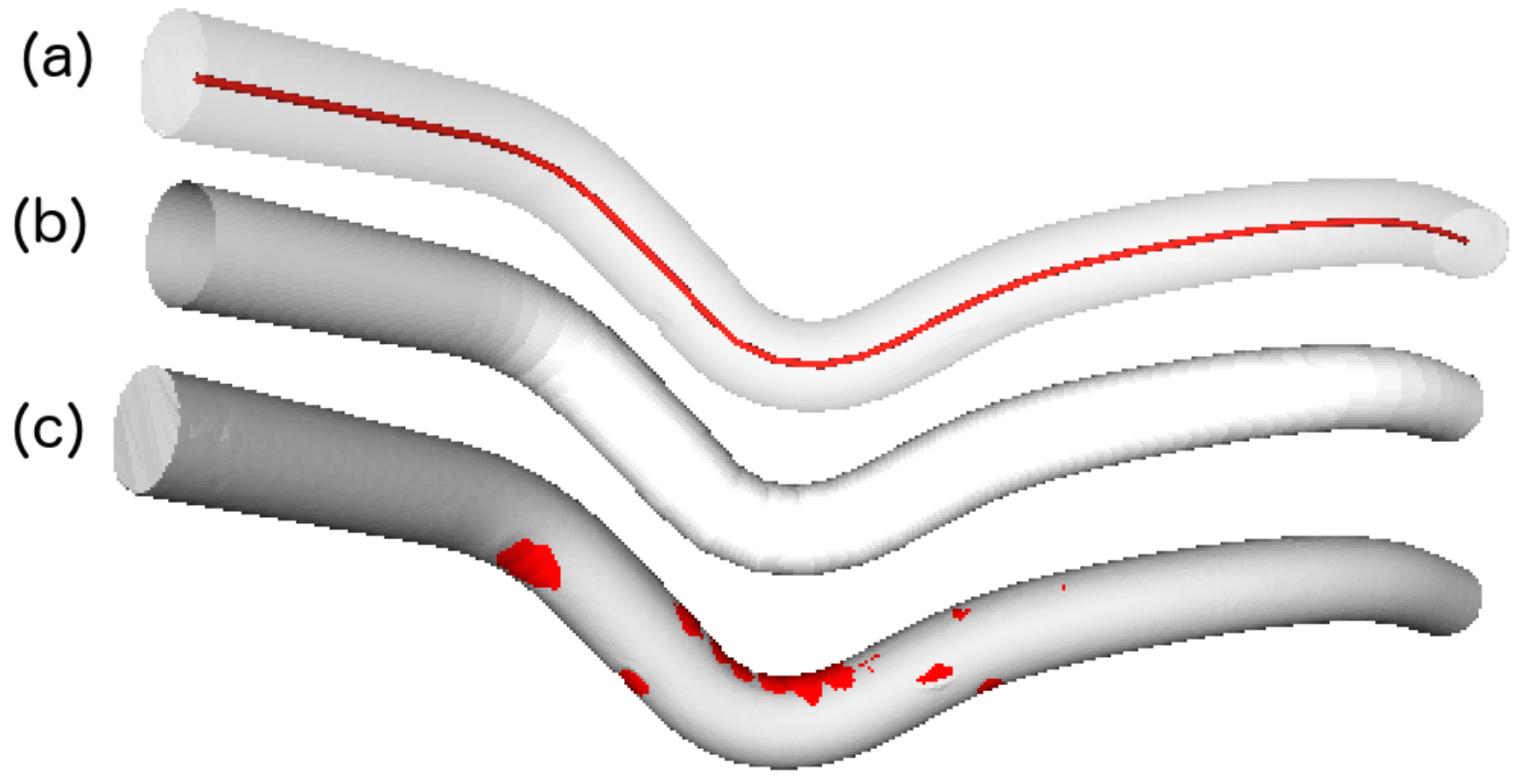}&
\includegraphics[width=0.401\textwidth]{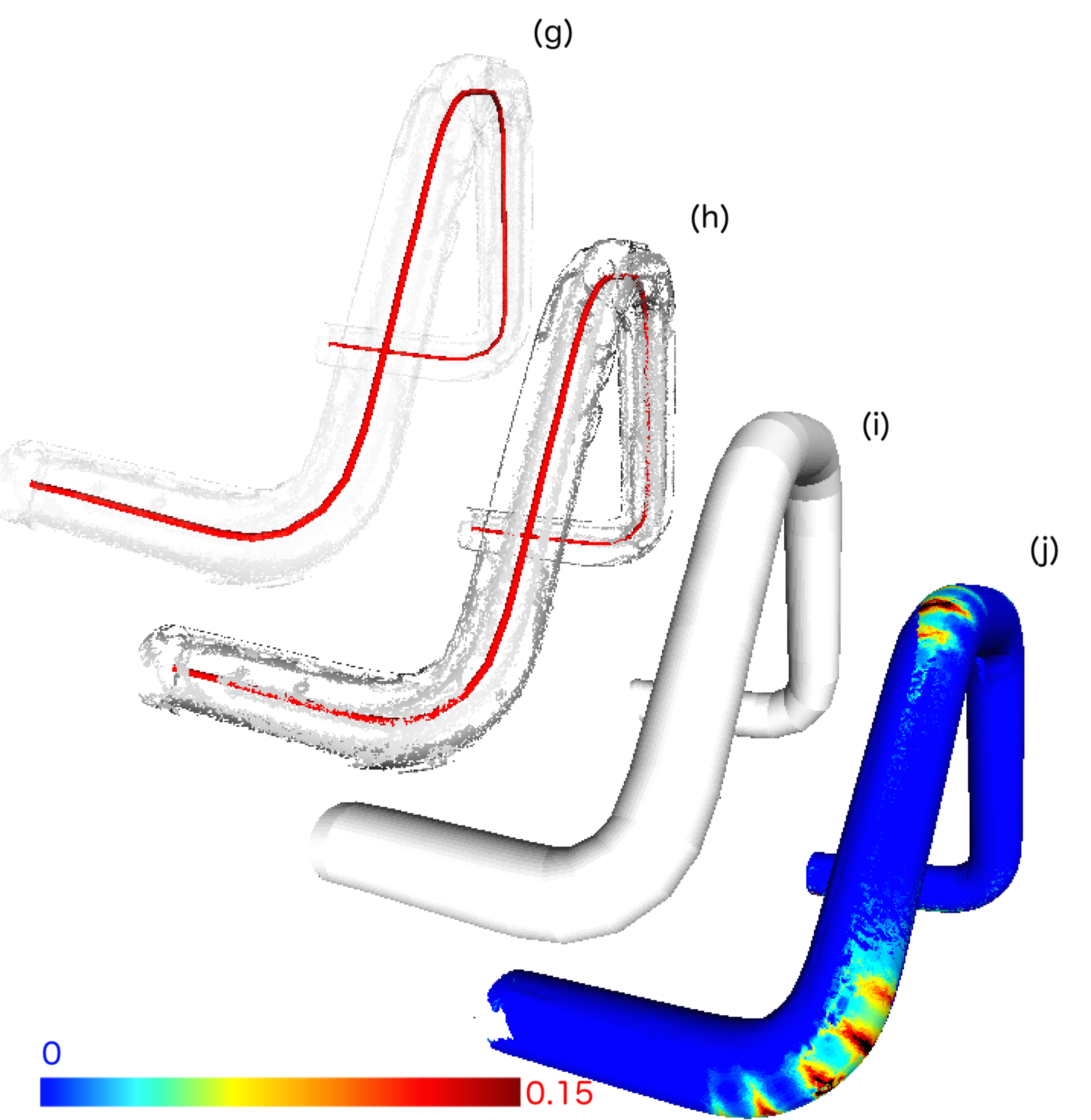}\\
\scriptsize Full mesh                       & \scriptsize Partial mesh\\[-.4em]
\scriptsize $R=6$, t=6.23s, 151 444 faces   & \scriptsize $R=6$, t=2.45s,  37 527 faces\\
\includegraphics[width=0.401\textwidth]{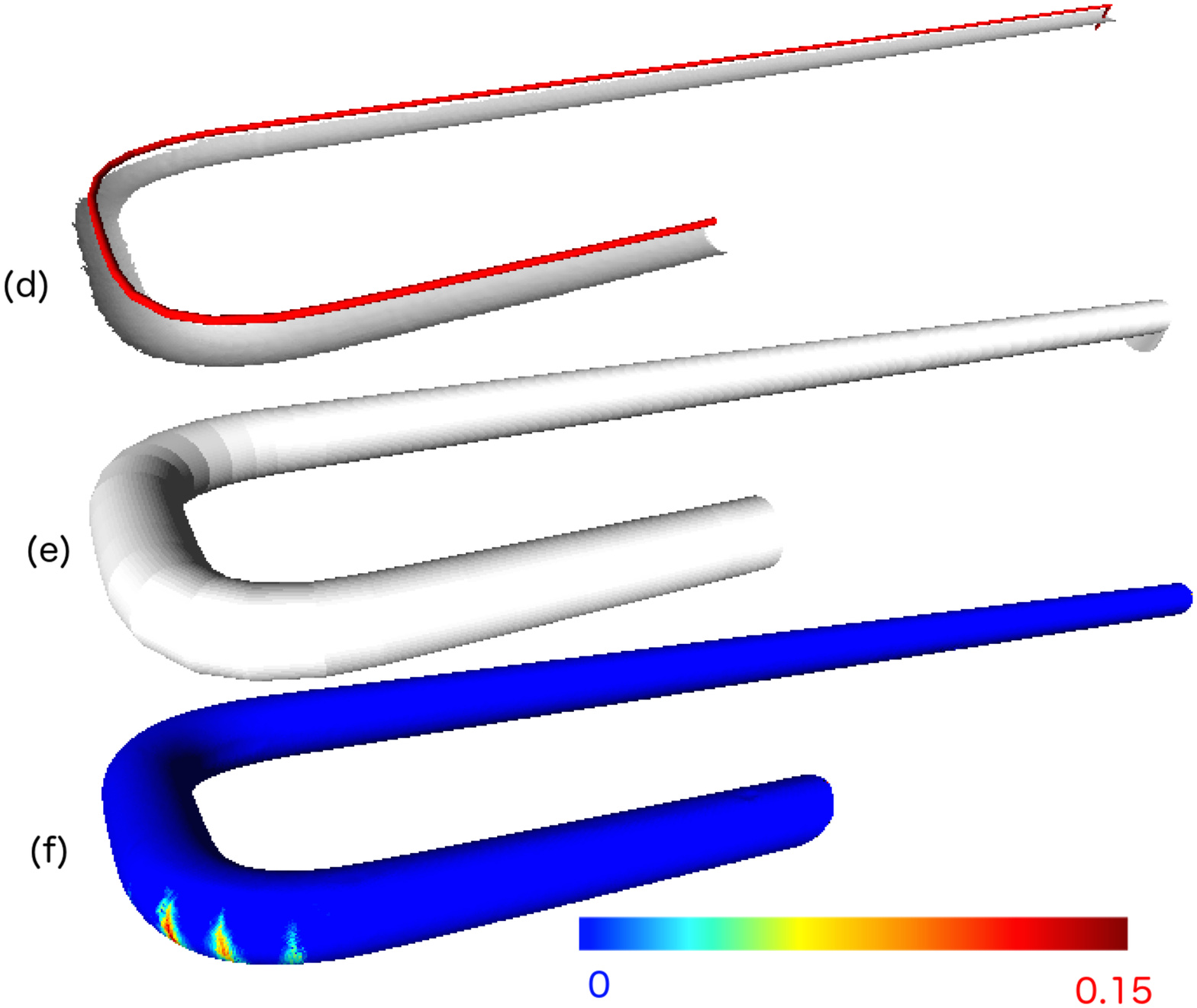}&
\includegraphics[width=0.401\textwidth]{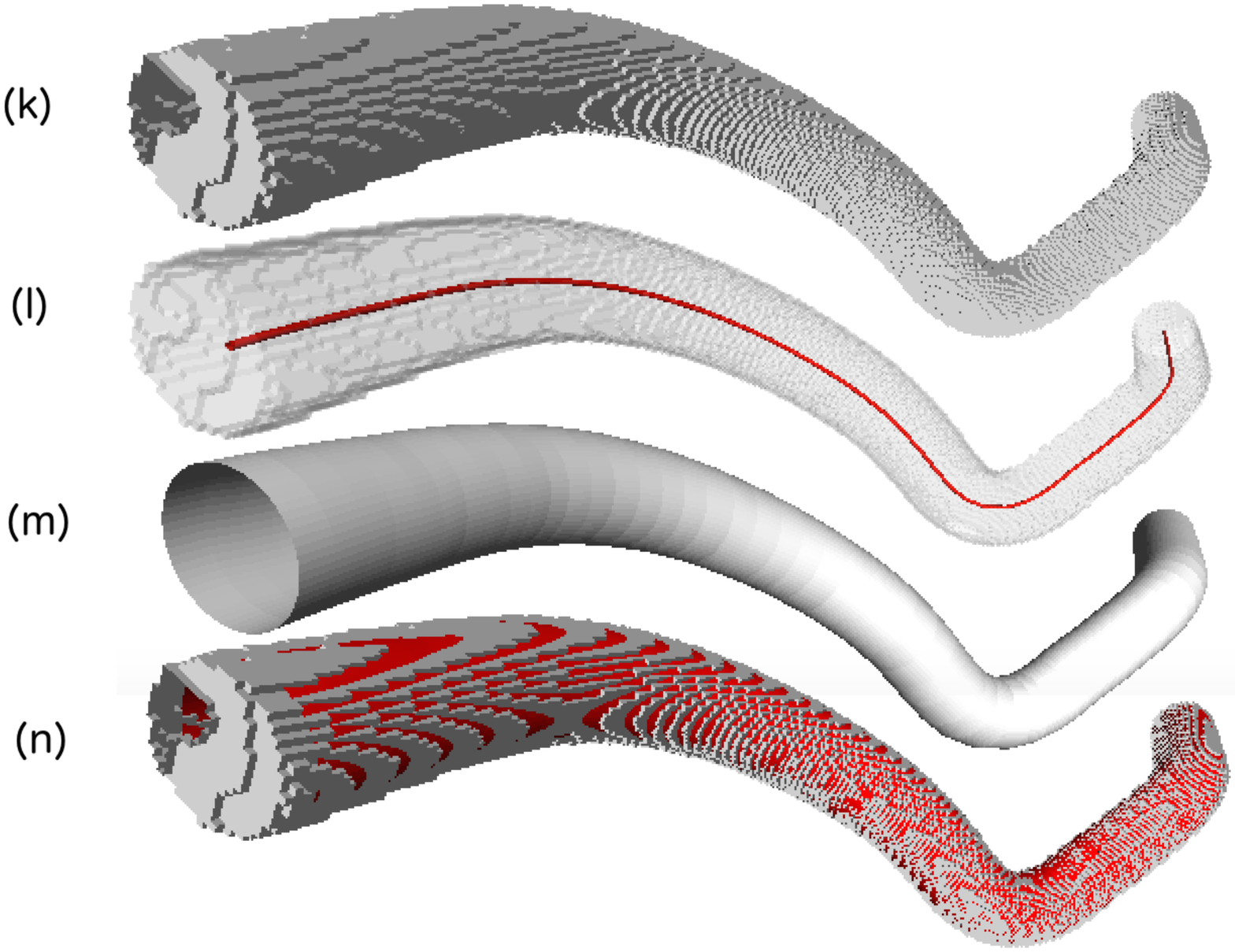}\\
\scriptsize Reduced scaned area             & \scriptsize Digital object\\[-.4em]
\scriptsize $R=6$, t=4.36s, 52 914 faces    & \scriptsize $R=4.9$, t=6.91s, 60 768 faces\\
\includegraphics[width=0.401\textwidth]{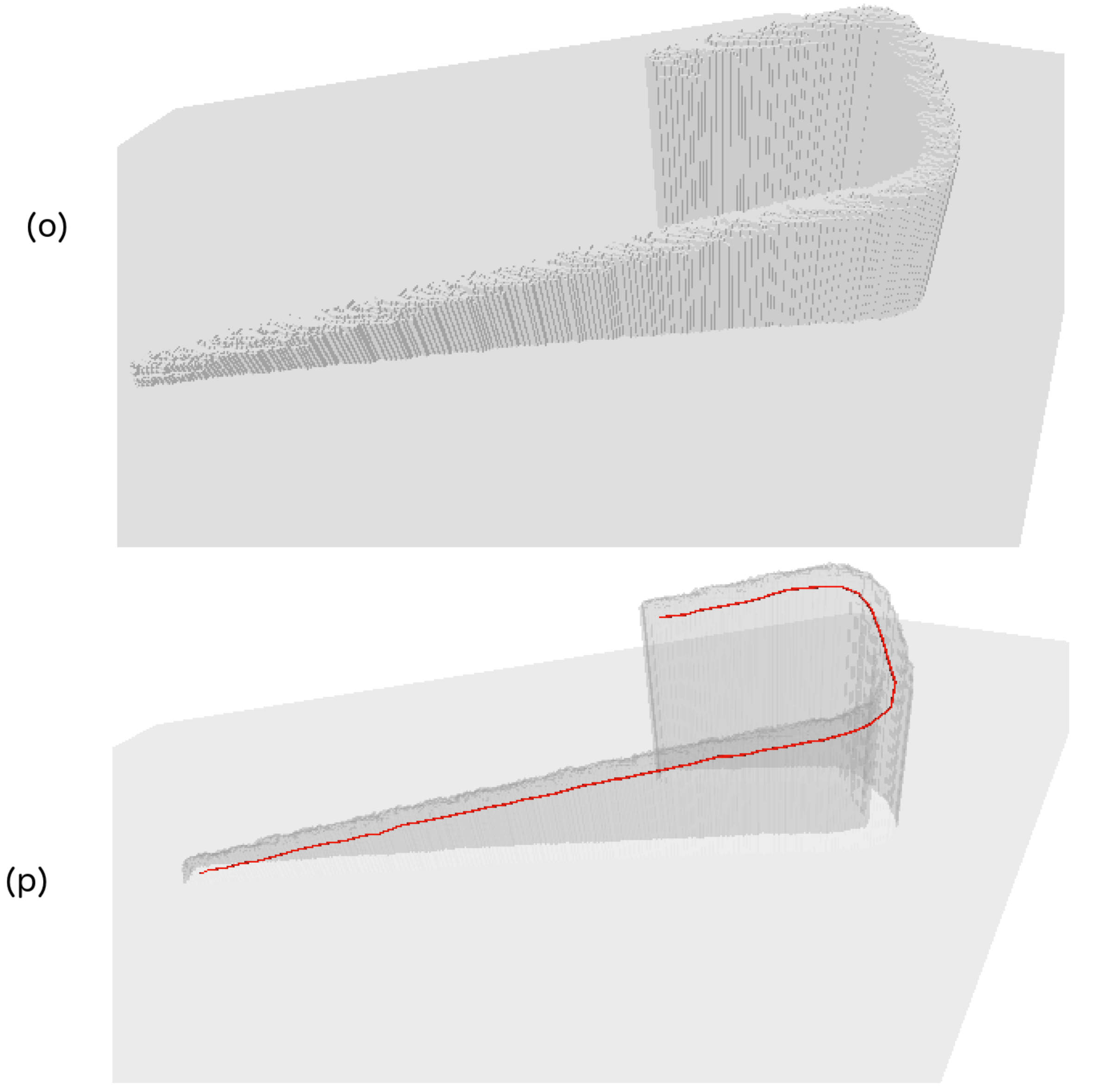}&
\includegraphics[width=0.401\textwidth]{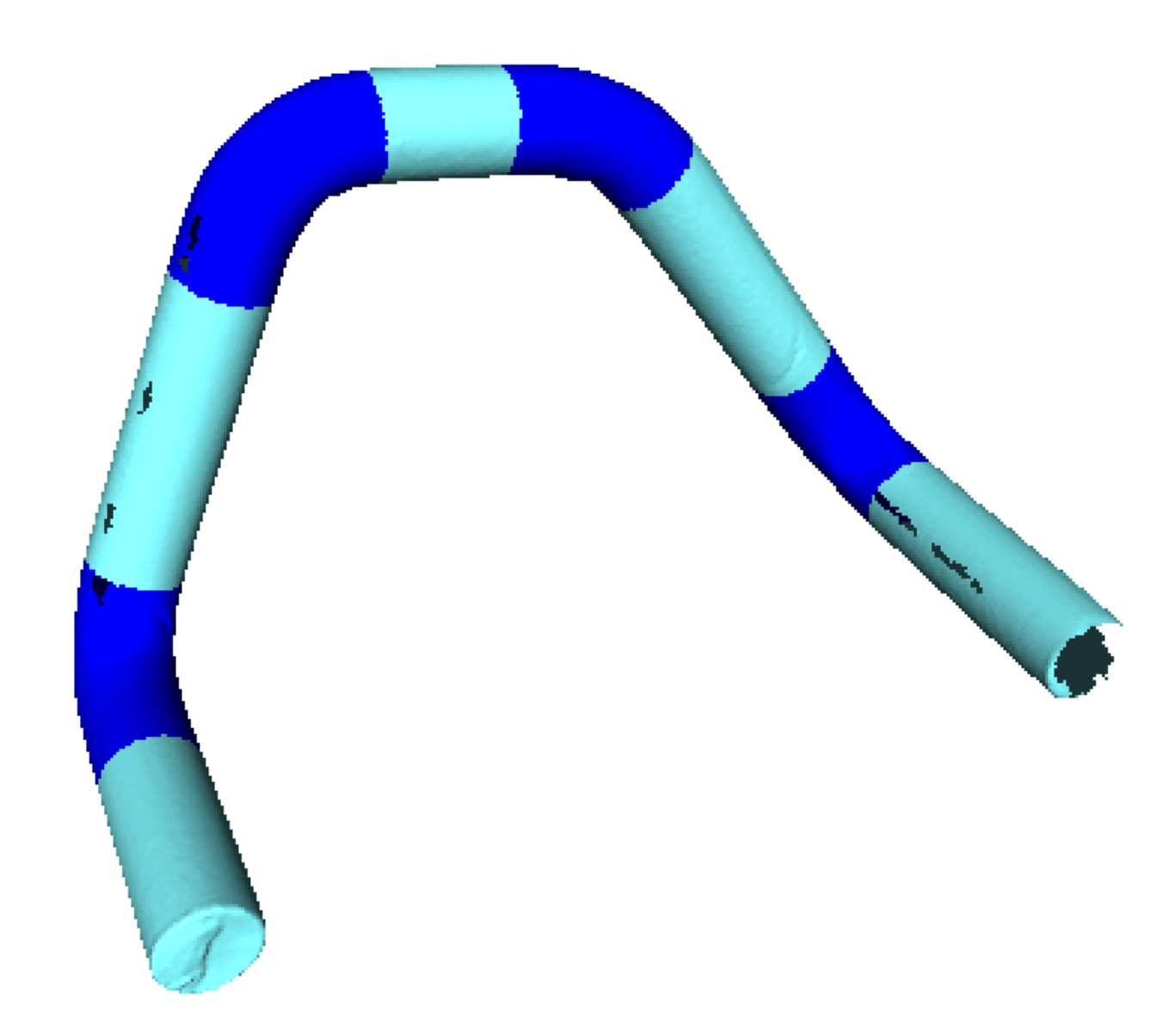}\\
\scriptsize Height map                      & \scriptsize (q) straight (in light blue) and toric \\[-.4em]
\scriptsize $R=8$,t=22.33s, 645 450 faces   & \scriptsize (in dark blue) segments. \\[-.4em]
                                            & \scriptsize $R=6$, t=12.17s, 187 638 faces \\[-.4em]
\end{tabular}
\caption{Result of reconstruction from various input data
  types. Images (a,g,l,p) show the centerline by transparency through
  the input surface, (k) and (o) are some input surfaces. (b,e,i,m)
  images are the reconstructed tubes built from the centerlines.
  Images (j) and (f) show the local squared distance error
  between the reconstructed tube and the input shape. Image (q)
  illustrates the decomposition of a tube into rectilinear and toric parts. For
  all experiments, the tracking parameter and epsilon were set
  respectively to $R$ and 0.001. Running times correspond to
  executions on a \textit{MacBook} computer with a 2,5 GHz \textit{Intel Core i7} processor.}
\label{FigReconsAll}
\end{figure}

{\bf Reconstruction results.} Several centerline extractions and
tubular shape reconstructions are shown on
\RefFigure{FigReconsAll}. Our input dataset contains several types of
metallic tubes numerized with different acquisition tools. In each
case, the centerline is always well delineated without the need to
tune a special parameter. When the input data is a complete or partial
mesh, the normal is simply estimated as the cross product of face
edges. When the input data is a digital object or a height map, we use
the digital Voronoi Covariance Measure \cite{cuel_vcm} to estimate the
normal vector. Parameters of this estimator are easily set since we
know the radius of the tubular object; they also have little influence
on the result, since the accumulation image makes the process very
robust. The running time was less than 30s for each experiment. As
show on image (j) and (f), few places present reconstruction
errors. Small errors may be found near bent areas. These errors are
less related to the reconstruction method than to the physical shape
under study, since the bending machine has deformed the tube at these
places.

~\\

\begin{floatingfigure}[r]{7cm}
\vskip -0.4cm
\hskip -0.6cm
\centering
\begin{tabular}{c@{\hspace{0.01\textwidth}}c}
\psfrag{c0}[c][][0.8]{\hskip 0.2cm\textbf{$C_0$}}
\psfrag{c1}[c][][0.8]{\hskip 0.2cm\textbf{$C_1$}}
\psfrag{c2}[c][][0.8]{\hskip 0.2cm\textbf{$C_2$}}
\psfrag{c3}[c][][0.8]{\hskip 0.2cm\textbf{$C_3$}}
\psfrag{c4}[c][][0.8]{\hskip 0.2cm\textbf{$C_4$}}
\psfrag{a1}[c][][0.8]{\hskip 0.2cm\textbf{$\alpha_1$}}
\psfrag{a2}[c][][0.8]{\hskip 0.2cm\textbf{$\alpha_2$}}
\psfrag{a3}[c][][0.8]{\hskip 0.2cm\textbf{$\alpha_3$}}

\includegraphics[width=0.28\textwidth]{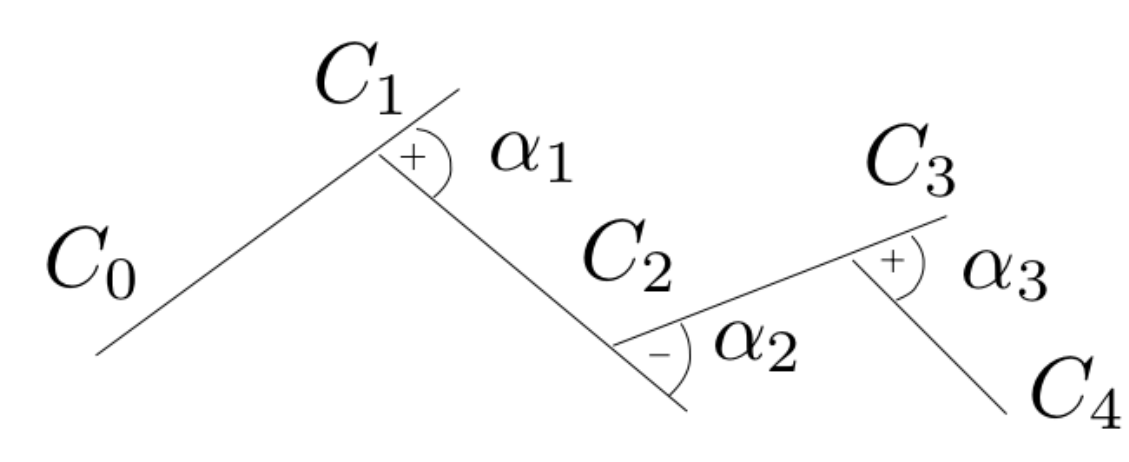}&
\psfrag{a1}[c][][0.8]{\hskip 0.2cm\textbf{$\alpha_1$}}
\psfrag{a2}[c][][0.8]{\hskip 0.2cm\textbf{$\alpha_2$}}
\psfrag{a3}[c][][0.8]{\hskip 0.2cm\textbf{$\alpha_3$}}

\psfrag{T02}[c][][0.7]{\hskip 0.2cm\textbf{$T_{02}$}}
\psfrag{T12}[c][][0.7]{\hskip 0.2cm\textbf{$T_{12}$}}
\psfrag{T11}[c][][0.7]{\hskip 0.2cm\textbf{$T_{11}$}}
\psfrag{T21}[c][][0.7]{\hskip 0.2cm\textbf{$T_{21}$}}
\psfrag{T22}[c][][0.7]{\hskip 0.2cm\textbf{$T_{22}$}}
\psfrag{T31}[c][][0.7]{\hskip 0.2cm\textbf{$T_{31}$}}
\psfrag{T32}[c][][0.7]{\hskip 0.2cm\textbf{$T_{32}$}}
\psfrag{T41}[c][][0.7]{\hskip 0.2cm\textbf{$T_{41}$}}

\includegraphics[width=0.29\textwidth]{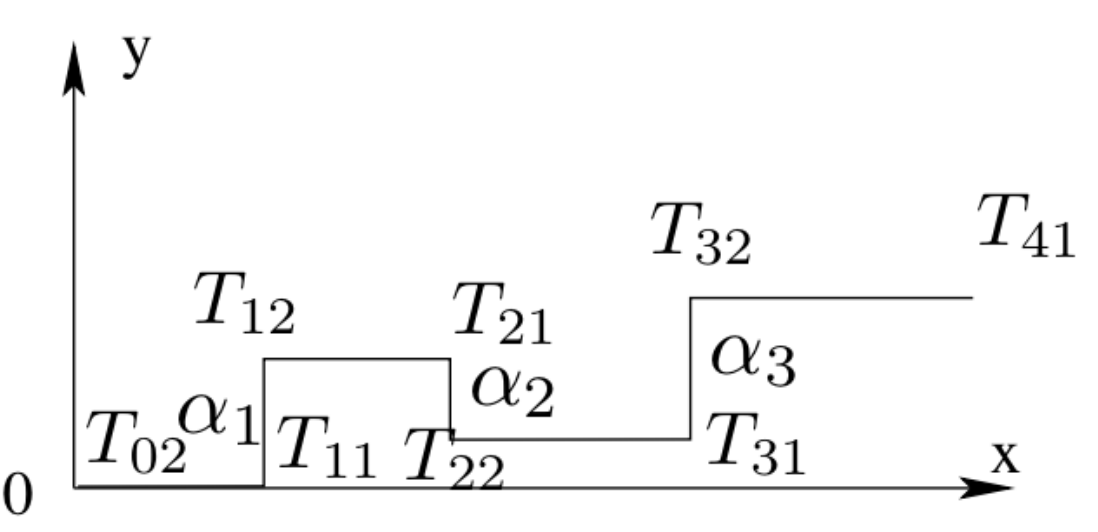}\\
\scriptsize{(a) input polygonal curve} &\scriptsize{\parbox{0.3\textwidth}{(b) Tangent space representation}}
\end{tabular}
\end{floatingfigure}

\noindent \textbf{Geometric analysis with 3D arc detection.}  We wish
to segment the tubular shape into rectilinear and toric parts. This
pro\-blem is equivalent to segmenting its centerline into 3D straight
segments and 3D circular arcs. We thus extend to 3D a method presented
by Nguyen {\em et al.} \cite{NguyenD11}, which was designed to cut a
2D discrete curve into straight and circular pieces. It relies on
properties of circular arcs in {\bf the tangent space representation}
that are inspired from Arkin~\cite{Arkin91} and
Latecki~\cite{Latecki00}. The tangent space representation of a
sequence of points $C=\{C_i\}_{i=0}^{n}$ is defined as follows :

Let $l_i$ be the length of segment $C_{i}C_{i+1}$ and $\alpha_i=\angle
(\overrightarrow {C_{i-1}C_i}, \overrightarrow {C_{i}C_{i+1}})$.  Let
us consider the transformation that associates $C$ with a polygon of
$\mathbb{R}^2$ constituted by segments $T_{i 2}T_{(i+1) 1},T_{(i+1)
  1}T_{(i+1) 2}$, $0 \leq i < n$ (upper floating figure) with:
$T_{02}=(0,0)$, $T_{i1}=(T_{(i-1)2}.x + l_{i-1}, T_{(i-1)2}.y)$ for
$i$ from $1$ to $n$, and $T_{i2}=(T_{i1}.x, T_{i1}.y+\alpha_i)$ for
$i$ from $1$ to $n-1$. %
Moreover, let $M=(M_i)_{i=0}^{n-1}$ be the sequence of midpoints of
segments $T_{i2}T_{(i+1)1}$ for $i$ from $0$ to $n-1$. The main idea
of the arc detection method is that if $C$ is a polygon that
approximates a circle or an arc of circle then $(M_i)_{i=0}^{n-1}$ is
a sequence of (approximately) co-linear points~\cite{NguyenD11}.

In our work, the sequence $(C_i)_i$ of skeleton points obtained in
Section~\ref{skeletonSection} is considered and the representation of
this sequence of points in the tangent space is computed. If the
angles $\alpha_k$, for $k$ from $p$ to $q$, of consecutive points of
$(C_k)_{k=p}^q$ are close to $0$, these points belong to a straight
line. Otherwise, the co-linearity of the corresponding midpoints
$(M_k)_{k=p}^q$ is tested in the tangent space by using an algorithm
presented in ~\cite{NguyenD11}. \RefFigure{FigReconsAll}~(q) shows an
example of tubular shape decomposition with this 3D variant of
circular arc detection. Toric and rectilinear parts are correctly
identified.

\section{Conclusion and Discussion}

A new efficient and simple method was presented to solve the problem
of delineating the centerline of 3D tubular shapes, for various types
of input data approximating its boundary: mesh, set of voxels or
height map. The method is robust to missing parts in input data as
well as perturbations: in these situations, it still returns
accurately the centerline position. To achieve this, we have
decomposed the process in three steps : 1) computing an accumulation
map from faces and their normal vectors, 2) tracking of centerline
through cross-section maximas of the accumulation map and 3)
optimization of the centerline position by a better fitting of the
model to the nearest faces along the centerline. The centerline was
accurate enough to allow further geometric analysis. We have shown how
to decompose the tubular shape into rectilinear and toric parts by a
simple adaptation of a 2D circular arc detection algorithm. The
hypothesis about constant radius parameter $R$ only influences the
skeleton position optimization. This limitation could be resolved
either by direct radius estimation from the accumulation image or by
radius optimization during position optimization. This is left for
future works.  The whole process was implemented with the
\textit{DGtal} \cite{dgtal} framework and will soon be available in
its companion \textit{DGtalTools}.

\section{Acknowledgments}
The authors would like to thank the anonymous reviewers and Nicolas
Passat for their many constructive comments, suggestions and references
that really helped to improve this paper.

\bibliographystyle{splncs03}
\bibliography{main}

\end{document}